\documentclass[namedreferences]{solarphysics}
\usepackage[optionalrh]{spr-sola-addons} 
\usepackage{graphicx}        
\usepackage{color}           
\usepackage{url}             

\begin{document}

\begin{article}

\begin{opening}

\title{On ponderomotive effects induced by Alfv\'en waves in inhomogeneous 2.5D MHD plasmas 	\\ {\it Solar Physics}}

\author{J.~O.~\surname{Thurgood}$^{1}$\sep
        J.~A.~\surname{McLaughlin}$^{1}$
       }
\runningauthor{Thurgood \& McLaughlin}
\runningtitle{Ponderomotive effects induced by Alfv\'en waves in 2.5D MHD}

   \institute{$^{1}$ Department of Mathematics \& Information Sciences, Northumbria University, Newcastle Upon Tyne,\\ NE1 8ST, UK\\
                     email: \url{jonathan.thurgood@northumbria.ac.uk}\\ 
             }

\begin{abstract}
Where spatial gradients in the amplitude of an Alfv\'en wave are non-zero, a nonlinear magnetic-pressure gradient acts upon the medium (commonly referred to as the \textit{ponderomotive force}). We investigate the nature  of such a force in inhomogeneous 2.5D MHD plasmas by analysing source terms in the nonlinear wave equations for the general case of inhomogeneous $\mathbf{B}$ and $\rho$, and consider supporting nonlinear numerical simulations. Our equations indicate there are two distinct classes of ponderomotive effect induced by Alfv\'en waves in general 2.5D MHD, each with \emph{both} a longitudinal and transverse manifestation; 
\\
i) Geometric Effects: Gradients in the pulse geometry relative to the background magnetic field cause the wave to sustain cospatial disturbances, the \emph{longitudinal}  and \emph{transverse daughter disturbances} - where we report on the transverse disturbance for the first time.
\\
ii) $\nabla(c_{A})$ Effects: Where a pulse propagates through an inhomogeneous region (where the non-zero gradients in the Alfv\'en-speed profile $c_{A}$ are non-zero), the nonlinear magnetic-pressure gradient acts to accelerate the plasma. Transverse gradients (phase mixing regions) excite independently propagating fast magnetoacoustic waves (generalising the result of Nakariakov et al.    1997, Solar Physics, 175, 93) and longitudinal gradients (longitudinally dispersive regions) perturb along the field (thus creating static disturbances in $\beta=0$, and slow waves in $\beta\neq0$).
\\ 
We additionally demonstrate that mode conversion due the nonlinear Lorentz force is a one-way process, and does not act as a mechanism to nonlinearly generate Alfv\'en waves due to propagating magnetoacoustic waves. We conclude that these ponderomotive effects are induced by an  Alfv\'en wave propagating in any MHD medium, and have the potential to have significant consequences on the dynamics of energy transport and aspects of dissipation provided the system is sufficiently nonlinear and inhomogeneous.
\end{abstract}
\end{opening}
\section{Introduction}
     \label{introduction} 
Due to the abundance of observational data confirming the existence and ubiquity of MHD wave motions in coronal plasma, it is clear that a well-developed theory of MHD wave propagation is required for understanding many dynamic processes ongoing in the coronal plasma (see reviews by, e.g.,  De Moortel \citeyear{ineke2005};  Nakariakov \& Verwichte \citeyear{NK2005}; Ruderman \& Erd\'elyi \citeyear{Ruderman2009};    Goossens et al.    \citeyear{goosens2011}; De Moortel \& Nakariakov \citeyear{ineke2012}).
The magnetic topology of the corona supports a wide variety of features, and as such can be highly inhomogeneous. Since the behaviour of MHD  waves is determined by the medium in which they propagate, an understanding of wave propagation in inhomogeneous media is essential. 

One important consequence of considering propagation in an inhomogeneous medium is the concept of \textit{phase mixing}. Simply, as a pulse or wave train propagates through the inhomogeneous medium, oscillations on neighbouring fieldlines become out of phase as local Alfv\'en speeds on each fieldline differ. This process naturally creates transverse gradients in the wave across the background magnetic field, and acts as a (linear) dissipative mechanism of Alfv\'en wave energy, and hence is often considered as a potential mechanism of coronal heating (see, e.g., Heyvaerts \& Priest \citeyear{HP83}, Browning \citeyear{Browning91}, Narain \& Ulmschneider \citeyear{NU1}; \citeyear{NU2}).  Typically, such studies have considered an inhomogeneity in density (to permit a unidirectional field, which makes mathematics more tractable), however as this paper will re-enforce, the process of phase mixing is strictly dependent on a non-uniform background Alfv\'en speed, i.e. is dependent on both density \underline{and} magnetic field profiles, and is not restricted to cases of density inhomogeneity only. 

There is a secondary consequence of these non-zero spatial gradients in the amplitude of a propagating MHD wave: nonlinear magnetic-pressure gradients and nonlinear magnetic tension (i.e. the nonlinear Lorentz force) act upon the plasma. 
In the MHD context, these forces are often referred to as the \emph{ponderomotive force}. Generally, a ponderomotive force is defined as a basic nonlinear force consisting of spatial gradients in a wave-field which has a non-vanishing effect when averaged over the period, however we note that the term has been used to refer to other forces (see Allan et al. \citeyear{Allan91} and Verwichte \citeyear{ErwinThesis} for discussions of the terms historical usage). 
The concept exists in areas of plasma physics other than MHD waves, in particular in the study of laser-plasma interaction where the force causes, including self-focusing (e.g. Chen \citeyear{Chen1984}) and channel formation (e.g. Boyd \& Sanderson \citeyear{boydsanderson}). The ponderomotive force in the MHD waves context (i.e. the nonlinear Lorentz force arising in fluid equations of motion) has been formally discussed by Dewar(\citeyear{Dewar70}) and Webb et al. (\citeyear{webb05pt2}).  For a thorough discussion of the ponderomotive force in various contexts, , and a demonstration of  the equivalence of particle-orbit ponderomotive force, see Allan et al. (\citeyear{Allan91}, section 3 and appendix A).

In our case (propagating MHD waves in inhomogeneous media), the effects of such a force are primarily of interest in that they have the potential to facilitate nonlinear mode conversion, the result of which has consequences on energy transport and dissipation.
In many previous studies of ponderomotive effects (e.g., in almost all of the forthcoming references), because such studies are concerned with phenomena due to propagating Alfv\'en waves, the authors use the term ponderomotive force to refer specifically to the nonlinear magnetic pressure force (as Alfv\'en waves do not act on the medium via nonlinear magnetic tension, e.g. as shown in $\S \ref{section:3}$). However, as reported in Thurgood \& McLaughlin (\citeyear{Me2012a}), propagating magnetoacoustic waves have a similar action on the medium, but in their case due to both nonlinear magnetic tension and nonlinear magnetic pressure. To avoid ambiguity, henceforth when referring to the force responsible for the investigated phenomena we avoid the term \lq{ponderomotive force}\rq{} in favour of the more specific \lq{nonlinear magnetic-pressure gradient}\rq{} or \lq{nonlinear magnetic tension}\rq{} as appropriate. We use the term \lq{ponderomotive effect}\rq{} to refer to the wider family of phenomena caused by the nonlinear Lorentz force of propagating MHD waves, and in this paper we primarily consider the specific set of ponderomotive effects induced by propagating Alfv\'en waves.
 Outside of the coronal plasma context, ponderomotive effects of MHD waves have been investigated in magnetospheric physics (e.g. Rankin et al.    \citeyear{Rankin94};
 \\ Tikhonchuck et al.    \citeyear{Tik95}; Allan \& Manuel \citeyear{AM96}) and as an acceleration mechanism in the solar wind (eg. Stark et al.    \citeyear{Stark95}). Within the context of coronal plasmas,  ponderomotive effects where considered initially by Hollweg (1971) and later by Nakariakov et al.    (\citeyear{NK97}), Verwichte (\citeyear{ErwinThesis}) and Verwichte et al.    (\citeyear{Erwin99}), the last three of which are key references in this paper. 

Verwichte (\citeyear{ErwinThesis}) and Verwichte et al.    (\citeyear{Erwin99}) considered the nonlinear evolution of an Alfv\'en wave in a homogeneous 1.5D model. They found that the nonlinear magnetic-pressure gradient was responsible for two features of the resultant wave behaviour. Firstly, they found that the propagating Alfv\'en wave sustains a disturbance which is longitudinal to the background magnetic field, and caused no net perturbation to the plasma (since the average nonlinear magnetic pressure over the wave period was zero). They dubbed this feature a \textit{ponderomotive wing}. Secondly, they found that the interaction between two crossing Alfv\'en waves gives rise to a non-zero net perturbation of plasma density. Hence, in warm plasmas, this \textit{cross-ponderomotive effect} was identified as a potential mode conversion mechanism, as would certain combinations of pulse geometry and background inhomogeneity leading to  non-zero longitudinal perturbations as the pulse propagates.

Nakariakov et al.    (\citeyear{NK97}) considered the nonlinear excitation of the fast wave by a propagating Alfv\'en wave which undergoes phase mixing in a unidirectional field structured by a background density profile (hence, an inhomogeneous background Alfv\'en speed). They derive governing wave equations, then evaluate them at an initial instant where only a pure, linear Alfv\'en wave is perturbing the system to find that the nonlinear fast wave equation contains a source term dependent on transverse gradients in the Alfv\'en wave amplitude, hence showing that nonlinear mode conversion is permitted in their scenario. This is then demonstrated in a numerical simulation, where an initial condition perturbs the medium to generate an Alfv\'en pulse with a profile $\varpropto\textrm{sech}^2\left(y\right)$ (where $\hat{\mathbf{B}}=\hat{\mathbf{y}}$). The pulse subsequently generates  fast magnetoacoustic waves as the simulation evolves. They conclude that Alfv\'en waves not only heat directly via the phase mixing dissipation method as per Heyvaerts \& Priest (\citeyear{HP83}) but also indirectly through nonlinear coupling to the fast wave, which is itself dissipative.

 Botha et al. (\citeyear{gert2000}) extended this work and found that the efficiency of the mode conversion is determined by the frequency and amplitude of the Alfv\'en wave, and by the gradient in the background Alfv\'en speed. Their work showed that the fast wave component eventually reached saturation, and concluded that (if the Alfv\'en wave amplitude amplitude is high enough) the nonlinear mode conversion can be a significant sink of Alfv\'en wave energy that otherwise would contribute to the linear phase mixing damping mechanisms as per  Heyvaerts \& Priest (\citeyear{HP83}). McLaughlin et al. (\citeyear{jamesphasemixing2011}) further extended the work into the visco-resistive case, where they found that the equilibrium density profile (and hence the location of heating) is significantly modified by the the visco-resistivity and the ponderomotive effects (i.e. drifting of the heating layer can occur).

In this paper, we focus on the possibility of the nonlinear Lorentz force as an agent for mode conversion in the general 2.5D MHD scenario, and also consider how ponderomotive effects identified in 1.5D models carry over. To do so, we first extend the source term analysis of Nakariakov et al.    (\citeyear{NK97}) to a MHD scenario which may be inhomogeneous in both magnetic induction and density (i.e. the background Alfv\'en speed is inhomogeneous). Then,  we emulate the numerical experiment of Nakariakov et al.    (\citeyear{NK97}) in light of our extended analysis to discuss previously unreported ponderomotive effects.

Thus, the paper is presented as follows; in $\S \ref{section:2}$ we derive wave equations for a general 2.5D cold plasma, to determine the set of nonlinear interactions permitted between the MHD wave modes. In $\S \ref{section:3}$, we analyse nonlinear source terms forced by general Alfv\'en waves, and further consider the case for harmonic Alfv\'en waves in $\S \ref{harmalfvenwave}$ \& $\ref{3.2}$ . In $\S \ref{section:simulations}$ we present results of the numerical experiment. Finally, we interpret our results and draw conclusions  in $\S \ref{section:conclusion}$.

\section{MHD wave equations in a  general, 2.5D $\beta=0$ medium} 
      \label{section:2}      
We first determine the role nonlinear terms of the MHD equations play in facilitating mode coupling in a ideal, 2.5D, $\beta=0$ plasma permeated by a general, potential, magnetic field, which we take as $\mathbf{B}_{0} = \left[B_x, B_y, 0 \right]$ such that $\nabla\times\mathbf{B}_{0}=\mathbf{0}$, with a background density $\rho_{0}=\rho_{0}(x,y)$. We set $\partial/\partial z=0$ throughout, i.e. take $\hat{\mathbf{z}}$ as the invariant direction. The governing, ideal, $\beta=0$ nonlinear MHD equations are:
\begin{eqnarray}
\rho\left[\frac{\partial\mathbf{v}}{\partial t}+(\mathbf{v}\cdot\mathbf{\nabla})\mathbf{v}\right]&=& \left(\frac{\mathbf{\nabla}\times\mathbf{B}}{\mu} \right)\times\mathbf{B}\quad \nonumber \\
\frac{\partial\mathbf{B}}{\partial t}&=&\mathbf{\nabla}\times(\mathbf{v}\times\mathbf{B})\quad\nonumber\\
\frac{\partial\rho}{\partial t}&=& -\mathbf{\nabla}\cdot(\rho\mathbf{v})  \quad
\label{MHDeqns}
\end{eqnarray}
where the standard MHD notation applies.
Initially, the medium is flow-free ($\mathbf{v}_{0}=\mathbf{0}$), and is perturbed by finite amplitude perturbations of the form $\mathbf{B}=\mathbf{B}_{0}+\mathbf{b}(x,y,z,t)$, $\rho=\rho_{0}+\rho_{1}(x,y,z,t)$, $\mathbf{v}=\mathbf{0}+\mathbf{v}_{1}(x,y,z,t)$. This yields:
$$\frac{\partial\mathbf{v}_{1}}{\partial t} - \frac{1}{\mu\rho_0}\left(\nabla\times\mathbf{b}\right)\times\mathbf{B}_{0}
=
\frac{1}{\mu\rho_0}\left(\nabla\times\mathbf{b}\right)\times\mathbf{b}-\frac{\rho_1}{\rho_0}\frac{\partial\mathbf{v}_{1}}{\partial t} - \left(1+\frac{\rho_1}{\rho_0}\right)\left(\mathbf{v}_{1}\cdot\nabla\right)\mathbf{v}_{1} \nonumber
$$
$$\frac{\partial\mathbf{b}}{\partial t}-\nabla\times\left(\mathbf{v}_{1}\times\mathbf{B}_{0}\right)=\nabla\times\left(\mathbf{v}_{1}\times\mathbf{b}\right)
$$
\begin{equation}
\frac{\partial\rho_1}{\partial t}+\nabla\cdot\left(\rho_{0}\mathbf{v}_{1}\right)=-\nabla\cdot\left(\rho_{1}\mathbf{v}_{1}\right)
\end{equation}
where the left-hand-side/right-hand-side governs the linear/nonlinear behaviour respectively.

We now decompose into $xyz$-components such that $\mathbf{v}_{1}=\left(v_x,v_y,v_z\right)$ and  $\mathbf{b}=\left(b_x,b_y,b_z\right)$, giving:
\begin{eqnarray}
\frac{\partial v_x}{\partial t}+\frac{B_y}{\mu\rho_0}\left(\frac{\partial b_y}{\partial x}-\frac{\partial b_x}{\partial y}\right) &=& N_1 \nonumber \\
\frac{\partial v_y}{\partial t}-\frac{B_x}{\mu\rho_0}\left(\frac{\partial b_y}{\partial x}-\frac{\partial b_x}{\partial y}\right) &=& N_2 \nonumber \\
\frac{\partial v_z}{\partial t}-\frac{1}{\mu\rho_0}\left(B_{x}\frac{\partial}{\partial x}+ B_{y}\frac{\partial}{\partial y}\right)b_z &=& N_3 \nonumber \\
\frac{\partial b_x}{\partial t}-\frac{\partial}{\partial y}\left(v_{x}B_{y}-v_{y}B_{x}\right)&=&N_4 \nonumber \\
\frac{\partial b_y}{\partial t}+\frac{\partial}{\partial x}\left(v_{x}B_{y}-v_{y}B_{x}\right)&=&N_5\nonumber \\
\frac{\partial b_z}{\partial t}-\left(B_{x}\frac{\partial}{\partial x}+B_{y}\frac{\partial}{\partial y}\right)v_{z}
&=&N_6 \nonumber \\
\frac{\partial\rho_1}{\partial t}+\frac{\partial}{\partial x}\left(\rho_{0}v_x\right)+\frac{\partial}{\partial y}\left(\rho_{0}v_y\right)&=&N_7
\end{eqnarray}
where the nonlinear components $N_{1},\dots,N_{7}$ are:
\begin{eqnarray}
N_1 &=& \frac{1}{\mu\rho_0}\left[\left(b_{y}\frac{\partial b_x}{\partial y}\right)-\left(b_{y}\frac{\partial b_y}{\partial x}+b_{z}\frac{\partial b_z}{\partial x}\right)\right] \nonumber \\
&-&\quad\frac{\rho_1}{\rho_0}\frac{\partial v_x}{\partial t} - \left(1+\frac{\rho_1}{\rho_0}\right)\left(v_{x}\frac{\partial}{\partial x}+v_{y}\frac{\partial}{\partial y}\right)v_x \nonumber \\
N_2 &=& \frac{1}{\mu\rho_0}\left[\left(b_{x}\frac{\partial b_y}{\partial x}\right)-\left(b_{x}\frac{\partial b_x}{\partial y}+b_{z}\frac{\partial b_z}{\partial y}\right)\right] \nonumber \\
&-& \quad\frac{\rho_1}{\rho_0}\frac{\partial v_y}{\partial t}- \left(1+\frac{\rho_1}{\rho_0}\right)\left(v_{x}\frac{\partial}{\partial x}+v_{y}\frac{\partial}{\partial y}\right)v_y \nonumber \\
N_3 &=& \frac{1}{\mu\rho_0}\left(b_{x}\frac{\partial}{\partial x}+b_{y}\frac{\partial}{\partial y}\right)b_{z}\nonumber\\
&-&\quad\frac{\rho_1}{\rho_0}\frac{\partial v_z}{\partial t}- \left(1+\frac{\rho_1}{\rho_0}\right)\left(v_{x}\frac{\partial}{\partial x}+v_{y}\frac{\partial}{\partial y}\right)v_z \nonumber \\
N_4 &=&\frac{\partial}{\partial y}\left(v_{x}b_{y}-v_{y}b_{x}\right) \nonumber\\
N_5 &=&-\frac{\partial}{\partial x}\left(v_{x}b_{y}-v_{y}b_{x}\right) \nonumber \\
N_6 &=&\frac{\partial}{\partial x}\left(v_{z}b_{x}-v_{x}b_{z}\right)-\frac{\partial}{\partial y}\left(v_{y}b_{z}-v_{z}b_{y}\right)\nonumber \\
N_7&=& -
\rho_{1}\left(\frac{\partial v_{x}}{\partial x}+\frac{\partial v_{y}}{\partial y}\right)
-\left(v_{x}\frac{\partial}{\partial x}+v_{y}\frac{\partial}{\partial y} \right)\rho_{1}
\nonumber
\end{eqnarray}
Note that the nonlinear Lorentz force in the momentum equation has been considered in the alternative tension-pressure form 
$$\left(\nabla\times\mathbf{b}\right)\times\mathbf{b}
=\left(\mathbf{b}\cdot\nabla\right)\mathbf{b}-\frac{1}{2}\nabla\left(\mathbf{b}\cdot\mathbf{b}\right)
$$
so that the contributions of the different aspects may be compared. Here,
\\$\left(\mathbf{b}\cdot\nabla\right)\mathbf{b}$ is the nonlinear magnetic tension and $\nabla\left(\mathbf{b}\cdot\mathbf{b}\right)/2$ is the nonlinear magnetic-pressure gradient. 

We now  seek wave equations governing the evolution of the permitted wave modes. To do so, we differentiate the momentum equation terms to link the system of equations. We then find the velocity components corresponding to the fast, slow (absent in $\beta=0$) and Alfv\'en wave by using the coordinate system of Thurgood \& McLaughlin (\citeyear{Me2012a}). 
They use this system to not only derive governing wave equations, but to allow the formation of initial conditions that correspond to pure, distinct wave modes and to decompose propagating waves into constituent modes (we utilise this approach in our simulations in $\S \ref{section:3}$).
In our 2.5D, Cartesian analysis the velocity components are thus $v_\perp =  \mathbf{v}_{1}\cdot\left(\hat{\mathbf{z}}\times\mathbf{B}_0\right) = -B_{y}v_{x}+B_{x}v_{y}$ (fast), $v_z$ (Alfv\'en), and $v_\parallel = \mathbf{v}_{1}\cdot\mathbf{B}_0 =B_{x}v_{x}+B_{y}v_{y}$ (longitudinal perturbations, where the slow mode is absent in $\beta=0$). Hence, the governing equations are:
\begin{equation}
\left[
\frac{\partial^{2}}{\partial t^2} - c_{A}^{2}\left(\frac{\partial^2}{\partial x^2} +\frac{\partial^2}{\partial y^2}\right)
\right]v_\perp
= 
B_{x}\frac{\partial N_2}{\partial t}-B_{y}\frac{\partial N_1}{\partial t}
+ c_{A}^{2}\left( \frac{\partial N_5}{\partial x}-\frac{\partial N_4}{\partial y}\right)
\label{fasteqn}
\end{equation}
\begin{eqnarray}
\left[
\frac{\partial^{2}}{\partial t^2} - \frac{1}{\mu\rho_{0}}
\left( B_{x}\frac{\partial}{\partial x}+B_{y}\frac{\partial}{\partial y}\right)^{2}
\right] &v_z&  \nonumber\\
=\frac{\partial N_3}{\partial t}
&+&\frac{1}{\mu\rho_0} \left( B_{x}\frac{\partial}{\partial x}+B_{y}\frac{\partial}{\partial y}\right)N_6 
\label{alfveneqn}
\end{eqnarray}
\begin{equation}
\frac{\partial^{2}v_\parallel}{\partial t^2} = B_{x}\frac{\partial N_1}{\partial t} + B_{y}\frac{\partial N_2}{\partial t}
\label{sloweqn}
\end{equation}
where $c_{A}=\sqrt{B_{0}^{2}/\mu\rho_{0}}$ is the background/equilibrium Alfv\'en speed that varies in the $xy$-plane, such that $c_{A}=c_{A}(x,y)$. 
Note that by setting the nonlinear terms of equations (\ref{fasteqn}-\ref{sloweqn}) to be zero (i.e. the right-hand-side) we revert to the linear regime, and can see that the linear, $\beta=0$ fast waves and Alfv\'en waves are completely decoupled, and there are no disturbances along the magnetic field.

\section{Source Term analysis} 
  \label{section:3}
  
 We now  consider conditions on the fluid variables that correspond to  a pure, linear Alfv\'en wave (as per Alfv\'en \citeyear{Alfven42}, i.e. waves driven only by magnetic tension) at an early time when other modes are taken to be absent (i.e. coupling has not yet occurred)  to  determine which terms of equations  (\ref{fasteqn}-\ref{sloweqn}) will  act as sources of other modes of oscillation in the general MHD case  (i.e. terms that contribute to the acceleration of velocity components that are initially zero in the absence of the corresponding wave mode), i.e.  we perform a source term analysis.
 
To do so, we take $v_z \neq 0$ and $b_z\neq0$ with $v_x=v_y=b_x=b_y=\rho_1=0$ (and so $v_\perp=v_\parallel=0$), as perturbations in the $\hat{\mathbf{z}}$-direction correspond linearly to a pure Alfv\'en wave (and the linear Alfv\'en wave does not perturb mass density). Note that this sets $N_{i}=0$ for $i=3,\dots,7$, and simplifies $N_1$ and $N_2$. The wave equations (\ref{fasteqn}-\ref{sloweqn}) become:
\begin{equation}
\frac{\partial^{2} v_\perp}{\partial t^2}= \frac{1}{\mu\rho_{0}}\frac{\partial}{\partial t}\left[b_z \left(B_{y}\frac{\partial}{\partial x}-B_{x}\frac{\partial}{\partial y}\right)b_z\right]
\label{eqn:nlfast_gen}
\end{equation}
\begin{equation}
\left[
\frac{\partial^{2}}{\partial t^2} - \frac{1}{\mu\rho_{0}}
\left( B_{x}\frac{\partial}{\partial x}+B_{y}\frac{\partial}{\partial y}\right)^{2}
\right] v_z
=
0
\label{fordisprel}
\end{equation}
\begin{equation}
\frac{\partial^{2}v_\parallel}{\partial t^2} = -\frac{1}{\mu\rho_{0}}\frac{\partial}{\partial t}\left[b_z \left(B_{x}\frac{\partial}{\partial x}+B_{y}\frac{\partial}{\partial y}\right)b_z\right]
\label{eqn:nllongtd_gen}
\end{equation}
again, the right-hand-side contains the nonlinear terms. 

Here, we see that there is a source term associated with $v_\perp$, thus in this case it is possible that $\partial^{2} v_\perp/\partial t^2\neq0$.  Inspection reveals that equation (\ref{eqn:nlfast_gen}) can be rewritten as:
\begin{equation}
\frac{\partial^{2} v_\perp}{\partial t^2}= -\frac{1}{\mu\rho_{0}}\frac{\partial}{\partial t}\left[ \hat{\mathbf{z}}\times\mathbf{B}_{0} \cdot \nabla\left(\frac{b_{z}^{2}}{2}\right)\right] 
\label{eq:Deriv_PMF_perp}
\end{equation}
\\
\emph{Hence, the Alfv\'en wave can cause a nonlinear magnetic-pressure gradient to arise which in turn causes the excitation of fast magnetoacoustic waves},
where gradients \textit{transverse} to $\mathbf{B}_{0}$ are responsible. Note that the term $\hat{\mathbf{z}}\times\mathbf{B}_{0} \cdot \nabla$ is the gradient in the direction across fieldlines. This concurs with the analysis of Nakariakov et al.    (\citeyear{NK97}), which considered nonlinear effects in a unidirectional field and found coupling terms associated with transverse gradients
(note that if we specifically consider equations \ref{fasteqn} and \ref{alfveneqn} for such a scenario, we recover their wave equations, see Appendix \ref{appendix:A}.)
 Hence, the  Alfv\'en wave exerts a transverse, nonlinear magnetic-pressure gradient in any situation in which it assumes non-zero gradients across the magnetic field. However as highlighted by Verwichte (\citeyear{ErwinThesis}) the net perturbation will only be non-zero if the ponderomotive force averaged over a wave-period is non-zero. 
 
 Similarly, we find that the Alfv\'en wave can cause a  nonlinear pressure gradient to arise which causes longitudinal perturbations to the equilibrium field. By rewriting (\ref{eqn:nllongtd_gen}) as 
\begin{equation}
\frac{\partial^{2}v_{\parallel}}{\partial t^2} = -\frac{1}{\mu\rho_{0}}\frac{\partial}{\partial t}\left[ \mathbf{B}_0 \cdot \nabla\left(\frac{b_{z}^{2}}{2}\right)\right] 
\label{eq:Deriv_PMF_par}
\end{equation}
we see that this is due to \emph{longitudinal gradients} in the Alfv\'en wave amplitude, i.e. this is the longitudinal effect of nonlinear magnetic pressure.

We note that equations (\ref{eq:Deriv_PMF_perp}) and (\ref{eq:Deriv_PMF_par}) can be integrated with respect to time (see Appendix \ref{appendix:C}) to yield
\begin{equation}
\frac{\partial v_\perp}{\partial t}=-\frac{\hat{\mathbf{z}}\times\mathbf{B}_{0}}{\mu\rho_0}\cdot\nabla\left(\frac{b_{z}^{2}}{2}\right)=-\frac{1}{\mu\rho_0}\nabla_{\perp}\left(\frac{b_{z}^2}{2}\right)
\label{eq:PMF_perp}
\end{equation}
\begin{equation}
\frac{\partial v_\parallel}{\partial t}=-\frac{\mathbf{B}_{0}}{\mu\rho_0}\cdot\nabla\left(\frac{b_{z}^{2}}{2}\right)=-\frac{1}{\mu\rho_0}\nabla_{\parallel}\left(\frac{b_{z}^2}{2}\right)
\label{eq:PMF_par}
\end{equation}
i.e., a familiar force equation (\lq{$\mathbf{F}=\mathrm{m}\mathbf{a}$}\rq{}) dependent on the square of the amplitude of the Alfv\'en wave field, conforming to the general definition of a \lq{ponderomotive force}\rq{} discussed in $\S 1$. Here, we have adopted the notation $\nabla_{\perp}\equiv\hat{\mathbf{z}}\times\mathbf{B}_{0} \cdot \nabla$ and $\nabla_{\parallel}\equiv \mathbf{B}_0 \cdot \nabla$ - these terms are the gradients transverse and longitudinal relative to the equilibrium magnetic field in the $xy$-plane.


One can also perform a source term analysis for fluid variables corresponding to an initially pure fast wave. Doing so, we find that the ponderomotive effects of the propagating fast wave do not facilitate coupling to the Alfv\'en mode, \emph{the fast wave does not interact with the Alfv\'en wave on any level, linear or nonlinear} in a medium with an invariant direction (see Appendix \ref{appendix:B}). Hence, we further can conclude that ponderomotive conversion in $\beta=0$ is a one-way process from the Alfv\'en to the fast magnetoacoustic mode.

\subsection{Harmonic Alfv\'en wave}
\label{harmalfvenwave}
To further investigate the nature of ponderomotive coupling, we reconsider the source terms of the fast wave and longitudinal equations when forced by a harmonic linear Alfv\'en wave of form
$$v_{z}=A\cos{\theta}=\Re\left(Ae^{i\theta}\right) \quad,\quad \theta=\omega t-\mathbf{k}\cdot\mathbf{r}\quad,\quad  b_{z}=S\left(x,y\right)v_{z}$$
where $S(x,y)$ is a spatial scaling term between velocity and magnetic field perturbations that is associated with the background Alfv\'en speed (for a linear Alfv\'en wave, $\mathbf{b}=\pm\sqrt{\mu\rho_{0}}\mathbf{v}$). If we consider that linear Alfv\'en waves propagate along magnetic field lines only, then the direction of the wavevector must be $\hat{\mathbf{k}}=\hat{\mathbf{B}}_0.$ We also can derive a dispersion relationship  from equation (\ref{fordisprel}, linear terms only) that links frequency, wave number and Alfv\'en speed:  $c_{A}^{2}=\omega^{2}/k^{2}$. Thus, we can consider the wavevector to be $\mathbf{k}=k\hat{\mathbf{k}}=k\hat{\mathbf{B}}_0=\omega\hat{\mathbf{B}}_0/c_{A}$, hence the function $\theta=\omega\left( t-\hat{\mathbf{B}}_{0}\cdot\mathbf{r}/c_{A}\right)$. 
Inserting the harmonic form of such a wave, the source terms driving the fast and longitudinal modes (equations (\ref{eqn:nlfast_gen} and \ref{eqn:nllongtd_gen}) become:
\begin{equation}
\frac{\partial^{2} v_\perp}{\partial t^2}=
\frac{A^{2}S^{2}}{\mu\rho_{0}}\frac{\omega^2}{c_{A}^{2}} 
\left[
\left(\hat{\mathbf{B}}_{0}\cdot\mathbf{r}\right)\nabla_{\perp}\left(c_{A}\right)
-c_{A}\nabla_{\perp}\left(\hat{\mathbf{B}}_{0}\cdot\mathbf{r}\right)
\right]
\cos{2\theta}\label{harmonicalven_driven_fast}
\end{equation}
\begin{equation}
\frac{\partial^{2}v_{\parallel}}{\partial t^2}=
\frac{A^{2}S^{2}}{\mu\rho_{0}}\frac{\omega^2}{c_{A}^{2}} 
\left[
\left(\hat{\mathbf{B}}_{0}\cdot\mathbf{r}\right)\nabla_{\parallel}\left(c_{A}\right)
-c_{A}\nabla_{\parallel}\left(\hat{\mathbf{B}}_{0}\cdot\mathbf{r}\right)
\right]
\cos{2\theta}\label{harmonicalven_driven_longt}
\end{equation}
%
Hence, equations (\ref{harmonicalven_driven_fast}) and (\ref{harmonicalven_driven_longt}) govern the transverse and longitudinal perturbations forced via the action of a harmonic Alfv\'en wave. 

The forced motion is driven at double the frequency ($\cos{2\theta}$) and is of order $\mathcal{O}(A^2)$, i.e. key features of the ponderomotive effect previously reported in the literature (see e.g., Nakariakov et al.    \citeyear{NK97}, Botha et al.    \citeyear{gert2000} and McLaughlin et al.    \citeyear{jamesphasemixing2011}). We find that perturbations to $v_\perp$ and $v_\parallel$ are dependent on two distinct classes of terms, one associated with gradients in the Alfv\'en-speed profile and the other with gradients in the geometry of the propagating pulse relative to the equilibrium magnetic field.

\subsection{Instantaneous, geometric terms: $\quad \nabla_{\perp}
\left(\hat{\mathbf{B}}_{0}\cdot\mathbf{r}\right)\quad$ \& $\quad \nabla_{\parallel}
\left(\hat{\mathbf{B}}_{0}\cdot\mathbf{r}\right)$}\label{3.2}
It is known  in 1.5D homogeneous MHD, for an Alfv\'en pulse propagating along a fieldline, that due to gradients in the wave intensity (i.e. ponderomotive effect) the leading flank acts to longitudinally-accelerate the plasma and the rear flank acts to longitudinally-decelerate the plasma, thus sustaining a \emph{cospatial, instantaneous  perturbation} that is transported along the magnetic field  which, depending on the specific pulse geometry, may  or may not cause a  non-zero (net) longitudinal perturbation as it passes through the medium (i.e. the \lq{ponderomotive wings}\rq{} of Verwichte \citeyear{Erwin99}). Regardless of whether a net perturbation to the plasma occurs, this \textit{longitudinal daughter disturbance} is continually sustained by the propagating Alfv\'en wave and always remains cospatial to its \textit{progenitor} (viz. the daughter occupies the same spatial region as the Alfv\'en wave). If we impose conditions equivalent to 1.5D homogeneous MHD (i.e. equations \ref{harmonicalven_driven_fast} and \ref{harmonicalven_driven_longt} where $\nabla_{\perp}=0$ and $\nabla_{\parallel}\left(c_{A}\right)=0$) only one driving term remains, 
$$\frac{\partial^{2}v_{\parallel}}{\partial t^2}\sim \nabla_{\parallel}\left(\hat{\mathbf{B}}_{0}\cdot\mathbf{r}\right)$$
i.e. this term  governs the longitudinal daughter disturbance.

Our equations indicate that, if we extend our consideration to the homogeneous  2.5D case (i.e. allow nonzero transverse gradients),  a transverse equivalent to Verwichte's ponderomotive wings, a \textit{transverse daughter disturbance}, can exist when there are transverse gradients in the pulse profile across the equilibrium magnetic field, i.e.: 
$$\frac{\partial^{2}v_{\perp}}{\partial t^2}\sim \nabla_{\perp}\left(\hat{\mathbf{B}}_{0}\cdot\mathbf{r}\right)$$
Such a profile, in homogeneous MHD, is imposed as an initial condition and would be maintained throughout, however in inhomogneous MHD a profile with transverse gradients can developss naturally via phase mixing, i.e where the pulse enters a region with a transverse \emph{Alfv\'en-speed profile}, thus assuming a pulse geometry with transverse gradients. 
We stress that such a transverse inhomogeneity of the Alfv\'en-speed profile (i.e. a phase mixing region) can be dependent on variable $\mathbf{B}_{0}$ and $\rho_0$, and is not just upon a density inhomogeneity in a unidirectional field.

\subsection{Inhomogeneity terms: $\quad \nabla_{\perp}\left(c_{A}\right)\quad$ \& $\quad \nabla_{\parallel}\left(c_{A}\right)$}\label{3.3}
Now permitting $\nabla(c_{A})\neq0$ and considering 2.5D inhomogeneous MHD, our equations show the ponderomotive effects not only depend upon the instantaneous magnetic pressure perturbations of assumed pulse geometry discussed previously,  but also upon magnetic pressure perturbations as the pulse geometry is altered from instant to instant (i.e. the full set of equations \ref{harmonicalven_driven_fast} and \ref{harmonicalven_driven_longt}).
This term is thus the ponderomotive effect due to Alfv\'en-speed profile inhomogeneities, where its transverse manifestation 
(which can be thought of as a  phase mixing term)
 excites fast waves, and the longitudinal manifestation excites  longitudinal perturbations (which can be thought of as a longitudinal dispersion term). If the net longitudinal perturbation is non-zero, due to the absence of gas pressure gradients in $\beta=0$ the perturbation will be static.

Thus, the analysis of $\S \ref{harmalfvenwave}$-$\ref{3.3}$ shows that the harmonic Alfv\'en wave nonlinearly interacts with the medium via the longitudinal and transverse manifestations of two classes of ponderomotive effect, namely the geometric effect (cospatial, ponderomotive daughter disturbances) and the  $\nabla(c_A)$ effect (ponderomotive acceleration due to inhomogeneous Alfv\'en-speed profile). Both may yield non-zero net perturbations of the medium, causing coupling to the longitudinal and transverse (fast magnetoacoustic) modes. As the two terms may interfere (constructively or destructively), the precise dynamics of ponderomotive mode conversion will vary on a case by case basis.


\section{Numerical Demonstration}\label{section:simulations}
We now demonstrate the ponderomotive effects identified in $\S \ref{section:3}$-$\ref{3.3}$ by considering simulations of a simple scenario
, namely that of an initially uniform pulse separating in a unidirectional field stratified by a smooth, yet steep, transverse density profile (i.e. a transverse Alfv\'en-speed profile). Such a scenario was considered in Nakariakov et al.    (\citeyear{NK97}), we emulate their results to demonstrate features that they did not report upon. 
 \begin{figure}    
    \centerline{\includegraphics[width=0.5\textwidth,clip=]{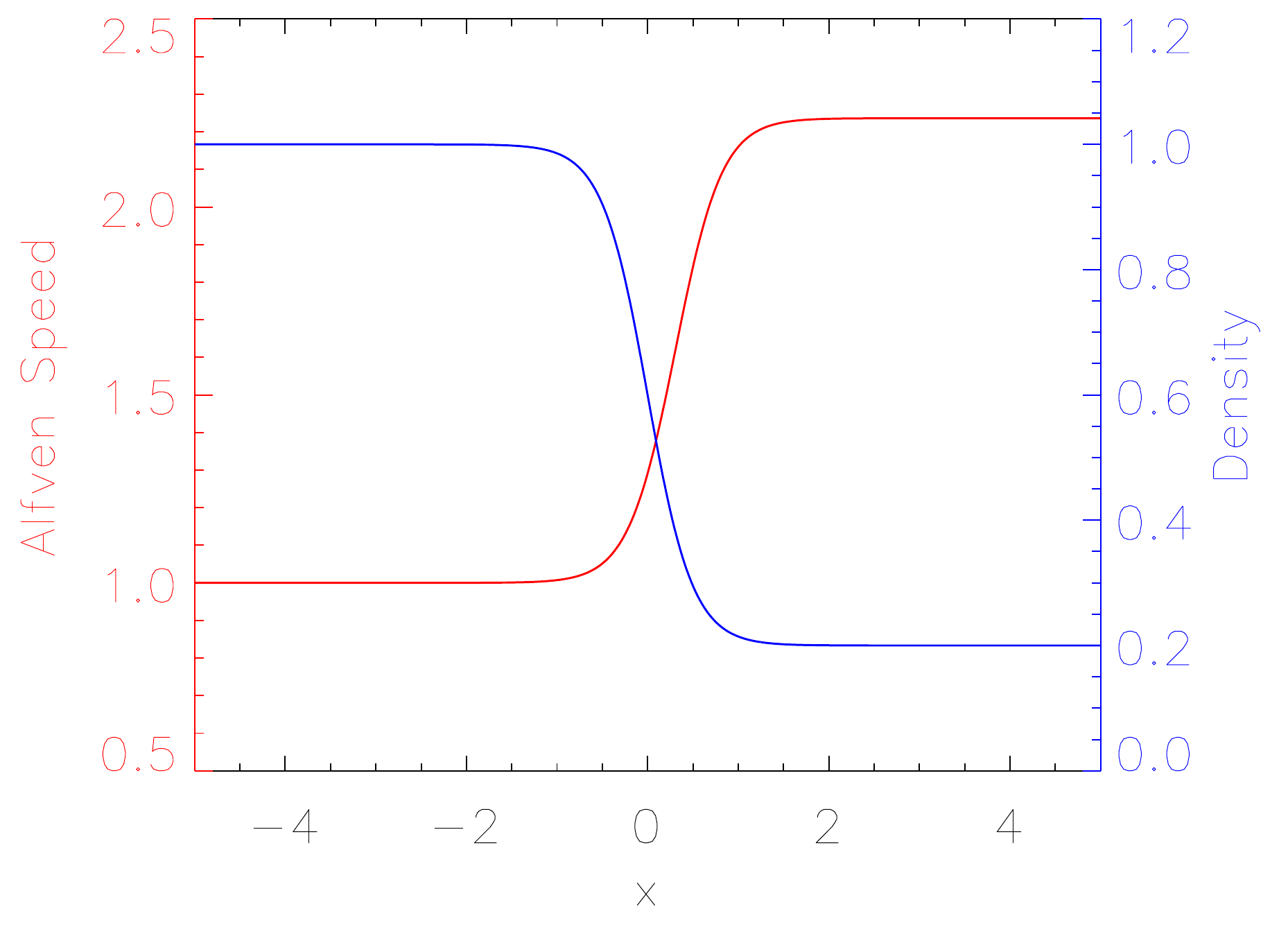}
              }   
\caption{The Alfv\'en-speed profile $c_A$ (red) and the density profile $\rho_0$ (blue). There is steep, continuous, transverse gradient in Alfv\'en speeds (a phase mixing region) in the region $-1<x<1$.}
   \label{a-s-p}
   \end{figure} 
We solve the nondimensionalised,  nonlinear MHD equations using LARE2D (see Arber et al.    \citeyear{LAREpaper} for details of code and, e.g., Thurgood \& McLaughlin \citeyear{Me2012a} for details of the nondimensionalisation procedure,  noting that all quantities are nondimensional in this section) for the equilibrium magnetic field $\mathbf{B}_{0}=\hat{\mathbf{y}}$ where the plasma is structured by a density inhomogeneity of the form:
$$
\rho_{0}=\rho_{0}\left(x\right)=
\frac{0.2+e^{-4 x}}{1.0+e^{-4x}}$$
in the numerical domain $x,y \in [-8,+8]$ with a resolution of $1920^2$ grid points for a cold plasma initially at rest ($\mathbf{v}_{0}=\mathbf{0}$).

 The resulting (nondimensionalised) Alfv\'en-speed profile, shown in Figure \ref{a-s-p}, is thus $c_{A}\approx1$ for $x < -1$, $c_{A}\approx \sqrt{5}$ for $x > +1$, with a steep but continuous transition between the two limits in the region $-1<x<1$. 
In this scenario (due to the unidirectional field) the longitudinal direction here is simply $\hat{\mathbf{y}}$ and the transverse direction is $\hat{\mathbf{x}}$. Thus the transverse and longitudinal velocity components of the simulation ($v_x$ and $v_y$) relates to the transverse and longitudinal velocity components  in the previous analysis ($v_\perp$ and $v_\parallel$) such that $v_\perp=-B_{y}v_{x}+B_{x}v_{y}=-v_{x}$ and $v_\parallel=B_{x}v_{x}+B_{y}v_{y}=v_{y}$.

\begin{figure}    
    \centerline{\includegraphics[width=1.0\textwidth,clip=]{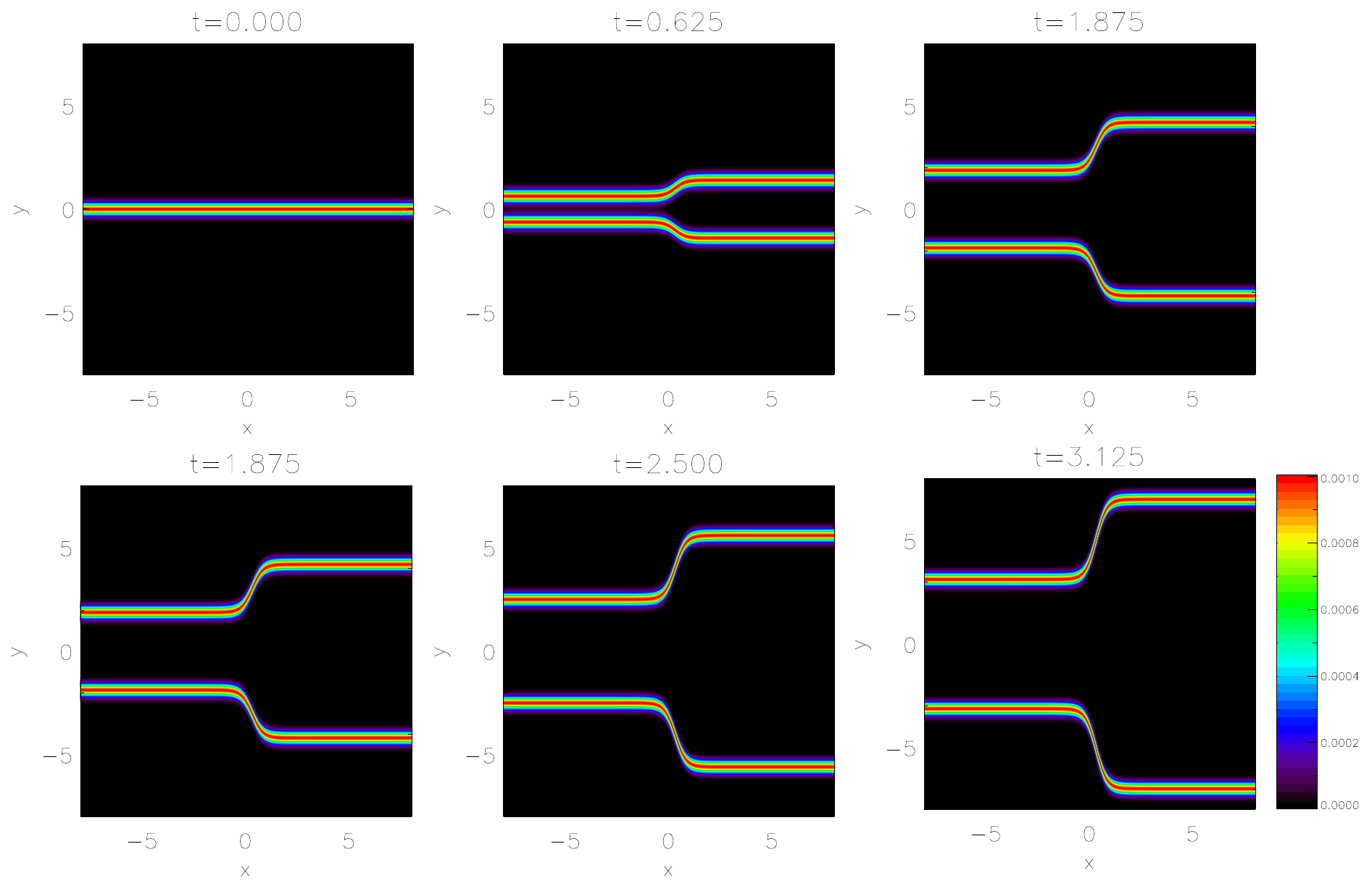}
              }   
\caption{Contour plots of $v_{z}$, showing the propagation of the Alfv\'en wave, which undergoes phase mixing due to the transverse inhomogeneity in Alfv\'en-speed profile. }
   \label{sim_vz}
   \end{figure}

We perturb the system by imposing an initial condition of the form
$$
v_{z}=2A 	\mathrm{sech}^2\left({\frac{y}{a}}\right) 
$$
with pulse-width parameter $a=0.25$, which creates an Alfv\'en wave (with apex at $y=0$) which separates into two oppositely-travelling pulses with amplitude $A=0.001$. 
The initial amplitude is taken to be small so that if a fast wave is nonlinearly generated, it will not disturb the equilibrium field through shock wave formation. 
Simple reflecting boundary conditions are employed, thus the simulation is halted once the pulses reach the boundaries.
Figure \ref{sim_vz} shows the resulting separation and propagation of the Alfv\'en waves (seen in $v_{z}$). Throughout time, the pulse geometry maintains its initial longitudinal ($\propto \mathrm{sech}^2y$) and transverse ($=0$) profiles outside of the phase mixing region ($x< -1$ and $x> +1$). In the region, due to phase mixing, the profiles change in time, with the creation of transverse gradients in pulse geometry which become increasingly steep as time progresses.

    \begin{figure}    
    \centerline{\includegraphics[width=1.0\textwidth,clip=]{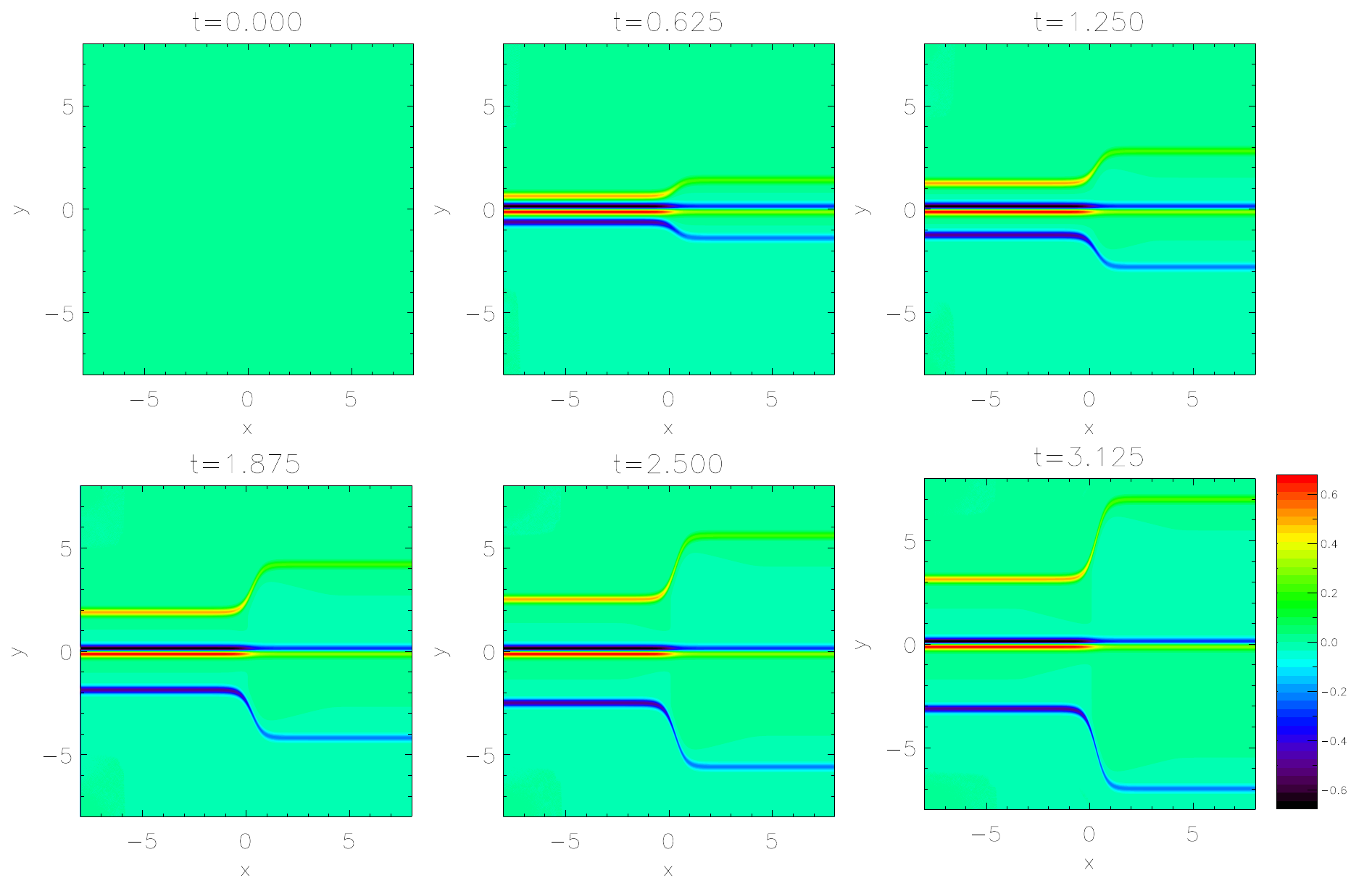}
              }
\caption{The longitudinal velocity component $v_{y}$ for the same times as Figure \ref{sim_vz}. We find a nonlinear static perturbation localised at the initial position of the Alfv\'en pulse and nonlinear cospatial disturbances that are transported with the Alfv\'en waves. The colour bar is scaled by a factor of $\times10^{6}$, i.e. by $\times A^{2}$.}
   \label{sim_longt}
   \end{figure}

  \begin{figure}    
    \centerline{\includegraphics[width=1.0\textwidth,clip=]{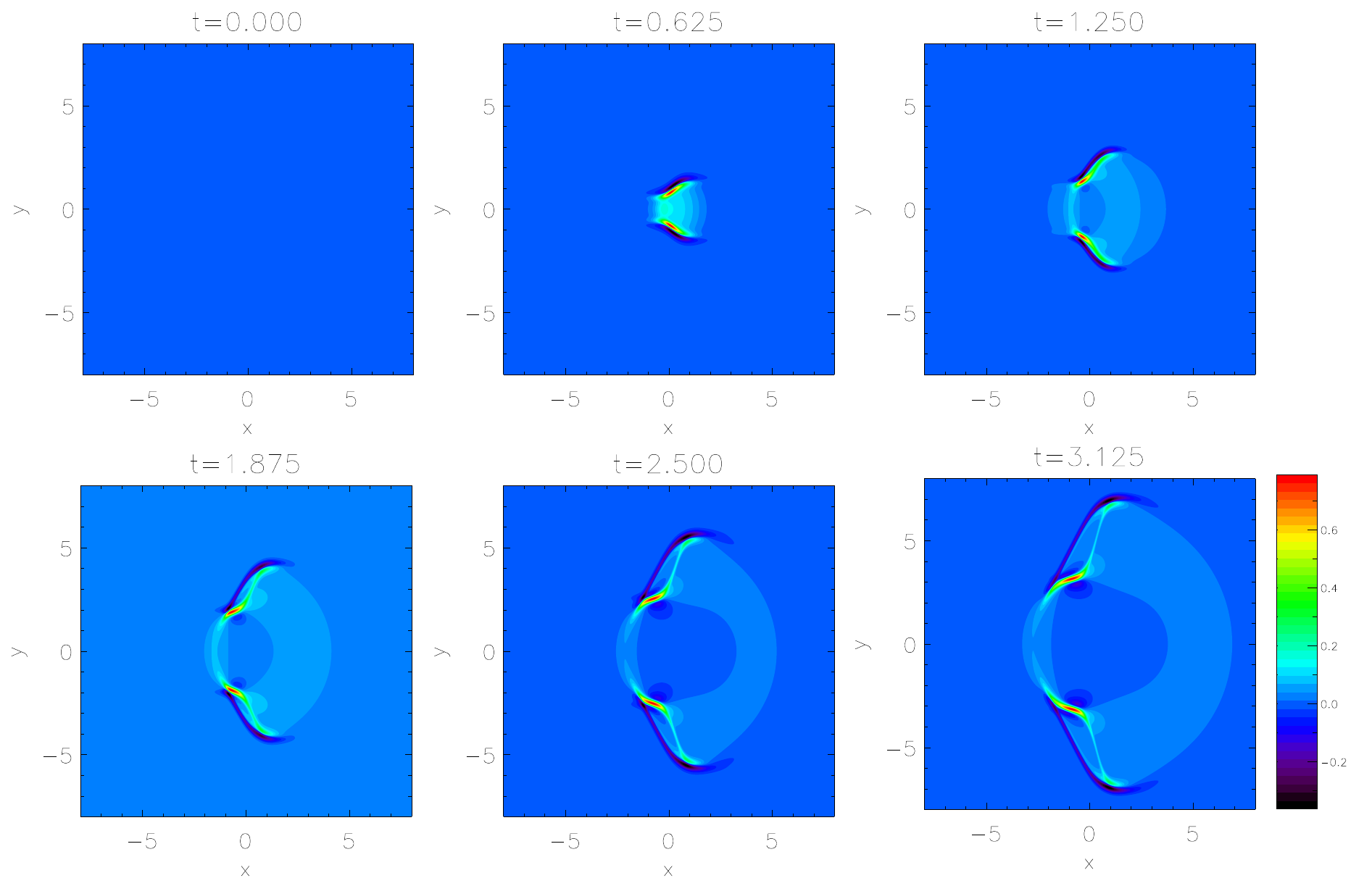}
              }    
\caption{The transverse velocity component $v_{x}$ for the same times as Figure \ref{sim_vz}. Propagating fast waves emanate from the phase mixing region. As the Alfv\'en wave undergoes phase mixing, independently propagating fast waves are generated and a cospatial transverse daughter disturbance develops. The colour bar is scaled by a factor of $\times10^{6}$, i.e. by $\times A^{2}$.}
   \label{sim_trans}
   \end{figure}      

We now consider perturbations to the longitudinal velocity component $v_{y}$, shown in Figure \ref{sim_longt}. Here, we note two nonlinear features. The first is a disturbance that appears to propagate as the Alfv\'en waves (i.e. along the fieldlines at the Alfv\'en speed) yet is observed in the longitudinal velocity component. These are not independently propagating waves (in our $\beta=0$ simulation there is no gas pressure to facilitate longitudinal oscillations) but  instantaneous perturbations/disturbances that are sustained and carried by the propagating Alfv\'en waves in $v_z$, i.e. these are the \textit{longitudinal daughter disturbances} identified in $\S 3.2$, and is associated with longitudinal gradients in the pulse geometry (for the harmonic Alfv\'en wave, this was of the form $\nabla_{\parallel}\hat{\mathbf{B}}_{0}\cdot\mathbf{r}$). The longitudinal daughter disturbance sustained by the upwardly propagating Alfv\'en wave is of the same sign, whereas the daughter sustained by the downwardly propagating wave is the opposite to its progenitor. Their maximum amplitudes are $v_{y}=\pm4.999\times10^{-7}$ respectively, which is of $\mathcal{O}\left(0.5A^{2}\right)$.  We note that amplitude of the longitudinal daughter is weaker by a factor of approximately $\sqrt{5}$ in the higher Alfv\'en speed region compared to its counterpart in the lower  Alfv\'en speed region


The second longitudinal ponderomotive effect is a static perturbation assuming the geometry of the initial pulse, which is also is of smaller amplitude in the higher Alfv\'en speed region. The fluid velocity is perturbed such that $v_{y}=\pm3.031\times10^{-7}$ at $x=-8$ and $v_{y}=\pm6.678\times10^{-8}$ at $x=+8$ (in the vicinity of $y=0$), hence this perturbation is also weaker in the higher Alfv\'en speed region by a factor of approximately $\sqrt{5}$, i.e. the amplitude of the longitudinal perturbations is inversely proportional to $c_{A}$. The perturbation is accompanied by a stationary mass density enhancement ($\rho_1$), initially generated at the same order as the velocity perturbations (the same magnitude and  with the inverse proportionality to $c_{A}$) then grows linearly in time during the simulation. These features are the 2.5D analogue of the cross-ponderomotive effect noted in the homogeneous 1.5D MHD study of Verwichte et al.    (\citeyear{Erwin99}), a non-zero longitudinal plasma perturbation caused by crossing pulses (which corresponds to our initial condition).

Figure \ref{sim_trans} shows the transverse velocity component $v_{x}$, which is the velocity component associated with the fast magnetoacoustic wave. As reported and thoroughly detailed in Nakariakov et al.    (\citeyear{NK97}), we find nonlinear disturbances in $v_{x}$ are generated in the phase mixing region and propagate outwards from regions of high-to-low Alfv\'en speed, independently of the Alfv\'en wave, clear evidence that the Alfv\'en wave nonlinearly generates independently propagating fast magnetoacoustic waves as it undergoes phasemixing. 
As the Alfv\'en wave is small, the fast waves saturates at a low amplitude ($v_{x}=-3.338\times10^{-7})$ relative to its progenitor, in agreement with Botha et al.    (\citeyear{gert2000}).
Aside from outwardly propagating fast waves of Nakariakov et al.    (\citeyear{NK97}), Figure \ref{sim_trans} reveals that as the Alfv\'en wave undergoes phase mixing, a transverse cospatial disturbance arises in the region, i.e. we can see the \emph{transverse daughter disturbance}. This is most clearly visible when comparing the later panels of Figure  \ref{sim_trans} to the corresponding times on Figure \ref{sim_vz} (to determine the region cospatial to the Alfv\'en wave).

\section{Conclusion}\label{section:conclusion}
In this paper, we present two main results:
\begin{itemize}
\item The analysis and conclusions of Nakariakov et al.    (\citeyear{NK97}) extend to the general  2.5D MHD case, such that the nonlinear magnetic pressure exerted by an Alfv\'en wave  propagating through a region of variable Alfv\'en speed facilitates mode conversion from the Alfv\'en to magnetoacoustic modes in many MHD scenarios (to the fast mode via phase mixing and to the longitudinal mode via dispersion along fieldlines).\\

\item The ponderomotive effect not only generates independently propagating magnetoacoustic waves but also longitudinal and transverse daughter disturbances, which remain cospatial and dependent upon the progenitor wave.

\end{itemize}

After deriving wave equations for a general MHD system (equations \ref{fasteqn}-\ref{sloweqn}) with a non-rotational invariance (i.e $\partial/\partial z=0$), the source term analysis of $\S \ref{section:3}$ demonstrates that the nonlinear magnetic-pressure gradients generated by a propagating Alfv\'en wave can facilitate  mode conversion from the Alfv\'en to magnetoacoustic modes, providing that, the pulse assumes a geometry which yields a non-zero net force over the wave's period. A similar source term analysis for a propagating fast wave is presented in Appendix \ref{appendix:B}, which confirms the additional result that the process is one-way
(i.e. the nonlinear Lorentz force does not permit conversion from the fast magnetoacoustic to the Alfv\'en mode).
 This, of course, is entirely intuitive as pressure gradients (magnetic or otherwise) cannot act in the Alfv\'en wave direction, as it is impossible to have gradients in an invariant direction (by definition).

In \S 3.1 we consider the terms forced by a harmonic Alfv\'en wave, and find that perturbations to $v_{\perp}$ and $v_\parallel$ are generated at double the frequency and at the square of the driving/initial  amplitude, in agreement with results reported in the literature. Such a result is necessary for ponderomotive effects, as the  nonlinear magnetic-pressure gradient acts upon the square of $b_{z}$ (and hence the square of $v_{z}$). The forced wave equations (\ref{harmonicalven_driven_fast}) and (\ref{harmonicalven_driven_longt}) are found to contain two distinct sets of terms, each with a longitudinal and transverse manifestation, that are:
\begin{itemize}
  \item  Geometric effects
  \begin{itemize}[(a)]
    \item $\quad \nabla_{\parallel}
\left(\hat{\mathbf{B}}_{0}\cdot\mathbf{r}\right)\quad$ - Longitudinal daughter disturbances  
\end{itemize}  
\begin{itemize}[(b)]
    \item $\quad \nabla_{\perp}
\left(\hat{\mathbf{B}}_{0}\cdot\mathbf{r}\right)\quad$ - Transverse daughter disturbances
  \end{itemize}
  \item $\nabla \left(c_{A}\right)$ effects
    \begin{itemize}[(c)]
    \item $\quad \nabla_{\parallel}\left(c_{A}\right)\quad$ - Perturbations due to longitudinal dispersion
    \end{itemize}  
\begin{itemize}[(d)]
    \item   $\quad \nabla_{\perp}\left(c_{A}\right)\quad$ - Perturbations due to phase mixing
  \end{itemize}
\end{itemize}

 Reducing to homogeneous 1.5D MHD, we find that one term forcing longitudinal perturbations remains, corresponding to $\nabla_{\parallel}
\left(\hat{\mathbf{B}}_{0}\cdot\mathbf{r}\right)$ above, which is dependent on longitudinal gradients in a function of pulse geometry and background magnetic field. This term is responsible for the cospatial \textit{ponderomotive wing} reported Verwichte et al.    (\citeyear{Erwin99}). 
Upon permitting a transverse direction (i.e. homogeneous 2.5D MHD), we find that a transverse analogue is permitted. Hence, to distinguish the two we refer to the longitudinal/transverse manifestations as the \textit{longitudinal daughter disturbance} and \textit{transverse daughter disturbance} respectively. When we further consider the inhomogeneous scenario and permit gradients in the Alfv\'en speed, a second class of forcing term is permitted, again with transverse and longitudinal manifestations, which accelerates the medium as the Alfv\'en wave passes through regions of inhomogeneity. The transverse manifestiation occurs where the Alfv\'en wave undergoes phase mixing (regions with transverse gradients in Alfv\'en speed) and the longitudinal manifestation occurs where longitudinal dispersion occurs (in regions with longitudinal gradients in Alfv\'en speed), where we stress that the Alfv\'en-speed profiles vary with $\mathbf{B}_{0}$ \underline{and} $\rho_0$ (i.e., phase mixing can occur in regions of uniform density, if the magnetic field has transverse stratification, an example of this is magnetic null point configurations, see e.g., Fruit \& Craig \citeyear{fruit_nullpointphasemixing}).

 To illustrate the phenomena implied by our analysis and interpretation of $\S \ref{section:3}$-$\S \ref{3.3}$, we considered numerical simulations of plasma with a unidirectional field structured by transverse density profile (the same scenario considered in Nakariakov et al.    \citeyear{NK97}).
In addition to the results detailed in Nakariakov et al.    (\citeyear{NK97}) - namely, the generation of independently propagating fast waves, we find clear evidence for the existence of the longitudinal daughter disturbance, which remains cospatial to its progenitor throughout its transit. We note that the disturbance is differs by a factor of $\sqrt{5}$ in the higher Alfv\'en speed region (relative to its amplitude in the lower Alfv\'en speed region). An explanation is found by reconsidering equation  (\ref{harmonicalven_driven_longt}) in the limit $x=-\infty$ and $x=+\infty$ for our scenario. Specifically, for the scenario there is no longitudinal dispersion ($\nabla_{\parallel}c_{A}=0$) and both the frequency and the longitudinal geometric gradient is identical at both extremes (thus we arbitrarily take $\omega=1$ and $\nabla_{\parallel}\left(\hat{\mathbf{B}}_{0}\cdot\mathbf{r}\right)=1$  without loss of generality), and the scaling function is taken as $S^{2}=\mu\rho_{0}$. The equation implies 
$$\textrm{max}\left(v_\parallel\right)
\approx
\frac{1}{2c_{A}}v_{z}^{2}
$$
Given that in the simulation $c_{A}\left(x=-\infty\right)=1$, $c_{A}\left(x=+\infty\right)=5$ with the Alfv\'en wave amplitude $A=0.001$  we find that
$v_\parallel\left(x=-\infty\right)=5.000\times10^{-7}$ and $v_\parallel\left(x=+\infty\right)=2.236\times10^{-8}$ which is in excellent agreement with the reported amplitudes of the longitudinal daughter disturbances in the homogeneous regions. Hence, the longitudinal daughter disturbances of Alfv\'en waves are generated with amplitude of $\mathcal{O}\left(A^{2}/2c_{A}\right)$.

The simulations also reveal the transverse manifestation of the cospatial disturbance, the transverse daughter. This is observed to develop over time (initially,  transverse  gradients in pulse geometry zero) as the Alfv\'en wave undergoes phase mixing, remaining cospatial to the phase mixed part of the Alfv\'en wave. To our knowledge, this is the first time this phenomena has been reported.

An additional effect noted in the simulation is that of the static perturbation at $y=0$ that is manifest in the  longitudinal, $\hat{\mathbf{y}}$-components of fluid variables (e.g. in $v_y$, Figure \ref{sim_longt}). We believe that this is the much the same cross-ponderomotive effect as reported in Verwichte et al.    (\citeyear{Erwin99}), despite the fact that our scenario is inhomogeneous 2.5D as opposed to the homogeneous 1.5D analysis they perform. This is because (a) the cross-ponderomotive effect will not have any transverse action as Alfv\'en waves cannot separate in the transverse direction (and there is nothing to suggest such a phenomena in the simulation data), and (b) there is no inhomogeneity along the field lines, and hence dispersion terms ($\nabla_{\parallel}c_{A}$) will not, in the specific case, impact upon the separation dynamics. 
As our simulation is inhomogeneous (with transverse stratification), it highlights that perturbations due to the cross-ponderomotive effect are (like the other ponderomotive effects observed) inversely  proportional to the speed $c_{A}$.  
It is however likely that in general, dispersion during pulse crossings will yield a more complicated effect. For cases where propagating Alfv\'en wave pulses are likely to meet (in particular where reflection may occur)
and the scenario is sufficiently nonlinear it would be necessary to investigate the ponderomotive effects on  interaction between Alfv\'en waves further. We suggest that it would be necessary integrate equation (\ref{harmonicalven_driven_longt}) over a separation period of a general wave form on $v_{z}$ (from which the form of $b_z$ follows) that corresponds to the general solution satisfying the variable-speed 1.5D wave equation $u_{tt}=c(y)u_{yy}$ for travelling pulses from a Gaussian-like initial condition (i.e. a variable speed D'Alembert solution). Such an analytic solution can be found by transforming the variable speed wave equation to a constant coefficient Klein-Gordon equation, however the profile of the wave speed has to conform to various conditions (of which, our scenario in $\S 4$ does not satisfy). See Grimshaw et al.    (\citeyear{grimshaw}) for a comprehensive overview of the homogenisation process.

During the separation of the pulse, we also note the generation of a stationary density enhancement (initially  of $\mathcal{O}\left(A^{2}/2c_{A}\right)$) that subsequently grows linearly in time. Again, this is in agreement with that reported by Verwichte et al. (\citeyear{Erwin99}). Linear density instabilities were shown to be a general feature of $\beta=0$ MHD by Falle and Hartquist (\citeyear{FH02}). There, the authors analyse the eigenmodes of cold MHD and show that there is a \emph{Jordan Mode Instability},  associated with the absent/zero-speed slow waves, which causes such density enhancements. This instability is to shown arise when the parallel (i.e. slow) velocity component is not constant. The cross-ponderomotive effect of a separating Alfv\'en wave (or two crossing Alfv\'en waves), as observed in our simulations, excites such a mode by acellerating the parallel velocity component by applying a (nonlinear) magnetic pressure gradient.
%


Throughout the paper we have considered the cold plasma regime. If the $\beta\neq0$ MHD wave equations (equations \ref{fasteqn}-\ref{sloweqn} if gas pressure gradients are permitted) are considered under the conditions for an initially pure Alfv\'en wave as per $\S \ref{section:3}$, we find that gas pressure gradients do not contribute, i.e. the $\beta\neq0$ equivalents of the source term derived equations (\ref{eqn:nlfast_gen}) - (\ref{harmonicalven_driven_longt}) are unchanged, i.e.  the way in which a passing Alfv\'en wave nonlinearly perturbs the medium is identical to that detailed in this paper. Hence, where non-zero longitudinal perturbations of the medium occur (which are static in $\beta=0$, e.g. the cross-ponderomotive perturbation in Figure \ref{sim_longt}), gas pressure gradients will subsequently arise to transport the disturbance longitudinally, i.e. slow magnetoacoustic waves will be generated. Thus, in $\beta=0$, ponderomotive mode conversion is permitted between the Alfv\'en wave and \emph{both} magnetoacoustic modes.
In addition to the implications for mode conversion in an extension to a $\beta\neq0$ regime, we note that the cross-ponderomotive effect, and subsequent excitation of a Jordan mode instability, will still generate large density enhancements in low-$\beta$ plasmas as per Falle and Hartquist (\citeyear{FH02}). Thus, crossing Alfv\'en waves will directly contribute to the creation of inhomogeneity in low-$\beta$ plasmas, due to their cross-ponderomotive effects.
%
%



We note that aspects of mode conversion, coupling and interaction, explored in this paper from the perspective of applied nonlinear forces, can be investigated using alternative methods, in particular a MHD instability approach. The relationship between the results presented here, and literature on Alfv\'en wave instabilities (such as, e.g., Derby \citeyear{Derby78}; Goldstein \citeyear{Goldstein78}; Champeaux et al. \citeyear{Champ97}; Webb et al. \citeyear{Webb01}) will be explored in future work.

We  conclude the paper by highlighting that all of the analysis can be repeated with different types of invariant coordinate system. For instance, we could consider azimuthal invariance, or the general coordinate system of Thurgood \& McLaughlin (\citeyear{Me2012a}) for zero-helicity topologies, and yield the same results. Given that an invariant coordinate is a necessary condition for the existence of true Alfv\'en waves as per Alfv\'en (\citeyear{Alfven42}) (see, e.g.,  Parker \citeyear{Parker91}	 and section on coordinate systems in Thurgood \& McLaughlin \citeyear{Me2012a}, their section 2.3.1), we believe our results and conclusions extend to any scenario in which an Alfv\'en wave (\emph{a wave driven by magnetic tension only}) can exist. 
We have shown that a nonlinear Lorentz force of a propagating Alfv\'en wave can generate independently propagating fast and slow magnetoacoustic waves, dependent on the interplay between Alf\'en speed profile and pulse geometry (viz., the form of the daughter disturbances), which varies on a case by case basis. As the transient properties of the magnetoacoustic modes are fundamentally different to the Alfv\'en wave, such conversion facilitates the indirect transport  and dissipation of (initially) Alfv\'en wave energy to plasma regions that are inaccessible in the  linear MHD regime. In modelling wave behaviour in sufficiently nonlinear solar plasmas, the ponderomotive effects of propagating waves must be evaluated.
\appendix

\section{Recovery of wave equations for a medium with a homogeneous magnetic field and a transverse density profile (Nakariakov et al.    1997)}\label{appendix:A}

Here we demonstrate that our general wave equations  for 2.5D MHD (\ref{fasteqn} \& \ref{alfveneqn}) reduce to those considered in the case of Nakariakov et al.    (\citeyear{NK97}), which considered a unidirectional homogeneous magnetic field structured by a transverse density profile. To do so, we take the field as constant in $\hat{\mathbf{y}}$, i.e  $\mathbf{B}_{0}=B_{0}\hat{\mathbf{y}}$, $B_{x}=0$ and $B_{y}=B_0$ and consider density as a function transverse to the field, i.e. $\rho_{0}=\rho_{0}(x)$. 
Thus, the Alfv\'en and fast wave equations (\ref{fasteqn} \& \ref{alfveneqn}) become

\begin{equation}
\left[\frac{\partial^2}{\partial t^2}-c_{A}^{2}(x)\frac{\partial^2}{\partial y^2}\right]v_z
= \frac{\partial N_3}{\partial t} +\frac{B_0}{\mu\rho_0(x)}\frac{\partial N_6}{\partial y}\nonumber \\
\label{Alfven_homoeqn}
\end{equation}
\begin{equation}
\left[
\frac{\partial^{2}}{\partial t^2} - c_{A}^{2}\left(\frac{\partial^2}{\partial x^2} +\frac{\partial^2}{\partial y^2}\right)
\right]v_\perp
=
-B_{0}\frac{\partial N_1}{\partial t}-c_{A}^{2}(x)\frac{\partial N_4}{\partial y}+c_{A}^{2}(x)\frac{\partial N_5}{\partial x}
\label{Fast_homoeqn}
\end{equation}

The Alfv\'en wave equation (\ref{Alfven_homoeqn}) is only superficially different from the equation governing the Alfv\'en wave in Nakariakov et al.    (\citeyear{NK97}), which (in their notation) is
\begin{equation}
\left[\frac{\partial^2}{\partial t^2}-c_{A}^{2}(x)\frac{\partial^2}{\partial z^2}\right]v_y
=\frac{1}{\rho_{0}(x)} \left(\frac{\partial N_2}{\partial t}+\frac{B_0}{4\pi}\frac{\partial N_5}{\partial z}\right)
\end{equation}
where the apparent difference is due to
\begin{itemize}
\item The analysis of Nakariakov et al.    (\citeyear{NK97}) considered $\hat{\mathbf{y}}$ as the invariant direction, hence $v_y$ corresponds to the Alfv\'en wave (as opposed to $v_z$ in this paper as $\hat{\mathbf{z}}$ is invariant). Additionally, $\partial / \partial y$ corresponds to $\partial / \partial z$ in this paper and Nakariakov et al.    (\citeyear{NK97}) respectively.
\item Hence, the nonlinear terms have a different ordering; their $N_1$, $N_2$, $N_3$, $N_4$, $N_5$, $N_6$ ,$N_7$ corresponds to our $N_1$, $N_3$, $N_2$, $N_4$, $N_6$, $N_5$, $N_7$ respectively.
\item The nonlinear terms originating in the equation of motion ($N_1$, $N_2$, $N_3$) are not exactly identical. They differ by a factor of $\rho_{0}^{-1}$ (we divided through the equation of motion by $\rho_0$ to group it with the nonlinear terms, they left it with the linear terms, e.g., our $N_1$ is equal to their $N_{1}/\rho_{0}$).
\item Nakariakov et al.    (\citeyear{NK97}) use cgs units, as opposed to SI in this paper.
\end{itemize}

 The fast wave equation of Nakariakov et al.    (\citeyear{NK97}) is 
\begin{equation}
\left[\frac{\partial^2}{\partial t^2}-c_{A}^{2}(x)
\left(\frac{\partial^2}{\partial x^2} +\frac{\partial^2}{\partial y^2}\right)
\right]v_x =
\frac{1}{\rho_{0}(x)} \left(\frac{\partial N_1}{\partial t}+\frac{B_0}{4\pi}\frac{\partial N_4}{\partial z} -\frac{B_0}{4\pi}\frac{\partial N_6}{\partial x} \right)
\end{equation}
the above only differs from our fast wave equation as per the aforementioned superficial differences. Additionally, the nonlinear terms (right hand side) differ by a factor of $-B_0$. The explanation is simple- in this case  $v_\perp=\mathbf{v}_{1}\cdot\hat{\mathbf{z}}\times\mathbf{B}_0=-B_{0}v_x$. Both correspond to a transverse perturbation of the field in this scenario and hence wave equations on either describe the evolution of the Alfv\'en wave; in a homogeneous field it is simpler to use $v_x$, for an inhomogeneous field $v_\perp$ is required.  Expressing $v_\perp$ in terms of $v_x$ and simplifying reduces our fast wave equation (\ref{Fast_homoeqn}) to the comparable form
\begin{equation}
\left[\frac{\partial^2}{\partial t^2}-c_{A}^{2}(x)
\left(\frac{\partial^2}{\partial x^2} +\frac{\partial^2}{\partial y^2}\right)
\right]v_x =
\frac{\partial N_1}{\partial t}+\frac{B_0}{\mu\rho_{0}(x)}\frac{\partial N_4}{\partial y}-\frac{B_0}{\mu\rho_{0}(x)}\frac{\partial N_5}{\partial x}
\end{equation}
%

\section{Nonlinear effects of a propagating fast wave.} 
  \label{appendix:B}

 We can  set $v_z=0$ and $b_z=0$ and place restrictions on   $v_x$, $v_y$, $b_x$, and $b_y$ that correspond to $v_{\parallel}=b_{\parallel}=0$,   $v_{\perp}\neq0$, $b_{\perp}\neq0$ and $\rho_1\neq0$ to consider the ponderomotive effects caused by a fast wave, yielding:
\begin{equation}
\left[
\frac{\partial^{2}}{\partial t^2} - c_{A}^{2}\left(\frac{\partial^2}{\partial x^2} +\frac{\partial^2}{\partial y^2}\right)
\right]v_\perp
= 
B_{x}\frac{\partial N_2}{\partial t}-B_{y}\frac{\partial N_1}{\partial t}
+ c_{A}^{2}\left( \frac{\partial N_5}{\partial x}-\frac{\partial N_4}{\partial y}\right)
\neq0
\label{fastselfinteraction}
\end{equation}
\begin{equation}
\frac{\partial^{2} v_z}{\partial t^2}= 0
\label{eqn:noMAtoAlfven}
\end{equation}
\begin{eqnarray}
\frac{\partial^{2} v_\parallel}{\partial t^2} &=& 
\frac{1}{\mu\rho_0}\frac{\partial}{\partial t}\left\lbrace B_{x}b_{y}\left[\left(\frac{\partial b_x}{\partial y}\right)_{T}-\left(\frac{\partial b_y}{\partial x}\right)_{P}\right] 
+B_{y}b_{x}\left[\left(\frac{\partial b_y}{\partial x}\right)_{T}-\left(\frac{\partial b_x}{\partial y}\right)_{P}\right]
\right\rbrace \nonumber \\
&=&\frac{1}{\mu\rho_0}\frac{\partial}{\partial t}
\left[
b_{\perp}
\left(\frac{\partial b_x}{\partial y}-\frac{\partial b_y}{\partial x}\right)
\right]
\label{fastinducedlong}
\end{eqnarray}
where subscripts $T$ and $P$ indicate whether a term is contributed by magnetic tension or magnetic pressure respectively, and $b_{\perp}=\mathbf{b}\cdot\hat{\mathbf{z}}\times\mathbf{B}_0$ is the transverse perturbation of the magnetic field.
The key result here is that \emph{equation (\ref{eqn:noMAtoAlfven}) shows that that the fast wave does not interact with the Alfv\'en wave on any level; linear or nonlinear}.

However, equation (\ref{fastinducedlong}) contains source terms, thus a propagating fast wave does cause nonlinear field-aligned disturbances that will cause, in certain circumstances, nonlinear Lorentz force coupling (via both nonlinear magnetic pressure/ponderomotive force and nonlinear magnetic tension
).
 Departing from the $\beta=0$ case, these source terms remain, i.e. the magnetoacoustic modes are coupled linearly by gas-pressure gradients, and nonlinearly by the Lorentz force. 

Equation (\ref{fastselfinteraction}) shows that such an initial fast wave undergoes self interaction. In this case, this is due to a combination of nonlinear Lorentz force, convective acceleration (contained within $N_1$ and $N_2$ which are permitted to be non-zero) and nonlinear induction ($N_4$ and $N_5$). Interestingly, the nonlinear induction term can be rewritten as
$$
c_{A}^{2}\left( \frac{\partial N_5}{\partial x}-\frac{\partial N_4}{\partial y}\right)=
-c_{A}^{2}\left( \frac{\partial^2 }{\partial x^2}+\frac{\partial^2}{\partial y^2}\right)
\mathbf{v}\cdot\hat{\mathbf{z}}\times\mathbf{b}
$$  
i.e. the nonlinear induction term is effectively of the same form as that of the linear motion due to velocity perturbations across the equilibrium field, however is instead motion due to perturbations across the induced magnetic field (though at the equilibrium, not induced, Alfv\'en speed).

Finally, we note that although there are no slow waves in the $\beta=0$ case, the nonlinear source term due to the Lorentz force acting upon the Alfv\'en wave (right hand side equation \ref{eqn:noMAtoAlfven}) would remain unchanged for $\beta\neq0$. This source term remains zero if an Alfv\'en wave is initially absent. Thus, the nonlinear Lorentz force does not facilitate slow to Alfv\'en mode conversion either. Hence, the analysis shows that for any 2.5D MHD scenario that that there is a one-way, nonlinear mode conversion mechanism between the Alfv\'en and (both) magnetoacoustic modes, that will become manifest when a phase mixed/ dispersive Alfv\'en wave (via  propagating in the vicinity of a inhomogeneous Alfv\'en-speed profile) assumes a geometry that exerts a non-zero average force over its period perturbing along and across the magnetic field.

\section{Derivation of force equations (\ref{eq:PMF_perp}) and (\ref{eq:PMF_par}).}   \label{appendix:C}
We demonstrate that equations (\ref{eq:Deriv_PMF_perp}) and (\ref{eq:Deriv_PMF_par}) can be integrated with respect to time to yield the force equations (\ref{eq:PMF_perp}) and (\ref{eq:PMF_par}) (i.e., with zero value \lq{integration constants}\rq{}) by simply considering an alternate, direct derivation from the momentum equation.

Under the source-term conditions of $\S \ref{section:3}$ ($v_z \neq 0$ and $b_z\neq0$ with $v_x=v_y=b_x=b_y=\rho_1=0$) the $\hat{\mathbf{x}}$ and $\hat{\mathbf{y}}$ components of the momentum equation simplify to yield 
$$
\frac{\partial v_{x}}{\partial t}
=-\frac{1}{\mu\rho_{0}}b_{z}\frac{\partial b_{z}}{\partial x}
=-\frac{1}{2\mu\rho_{0}}\frac{\partial b_{z}^{2}}{\partial x}
$$
$$
\frac{\partial v_{y}}{\partial t}
=-\frac{1}{\mu\rho_{0}}b_{z}\frac{\partial b_{z}}{\partial y}
=-\frac{1}{2\mu\rho_{0}}\frac{\partial b_{z}^{2}}{\partial y}
$$
From which equations (\ref{eq:PMF_perp}) and (\ref{eq:PMF_par}) readily follow by constructing equations in terms of $v_\perp$ and $v_\parallel$ from $v_\perp =  \mathbf{v}_{1}\cdot\left(\hat{\mathbf{z}}\times\mathbf{B}_0\right) = -B_{y}v_{x}+B_{x}v_{y}$ and $v_\parallel = \mathbf{v}_{1}\cdot\mathbf{B}_0 =B_{x}v_{x}+B_{y}v_{y}$.§


\begin{acks}
The authors acknowledge IDL support provided by STFC. JOT acknowledges travel support provided by the RAS and the IMA, and a Ph.D. scholarship provided by Northumbria University. The computational work for this paper was carried out on the joint STFC and SFC (SRIF) funded cluster at the University of St Andrews (Scotland, UK).
\end{acks}


\bibliographystyle{spr-mp-sola}

\bibliography{./references.bib}  

\begin{thebibliography}{34}
\ifx \bisbn   \undefined \def \bisbn  #1{ISBN #1}\fi
\ifx \binits  \undefined \def \binits#1{#1}\fi
\ifx \bauthor  \undefined \def \bauthor#1{#1}\fi
\ifx \batitle  \undefined \def \batitle#1{#1}\fi
\ifx \bjtitle  \undefined \def \bjtitle#1{\textit{#1}}\fi
\ifx \bvolume  \undefined \def \bvolume#1{\textbf{#1}}\fi
\ifx \byear  \undefined \def \byear#1{#1}\fi
\ifx \bissue  \undefined \def \bissue#1{#1}\fi
\ifx \bfpage  \undefined \def \bfpage#1{#1}\fi
\ifx \blpage  \undefined \def \blpage #1{#1}\fi
\ifx \burl  \undefined \def \burl#1{\textsf{#1}}\fi
\ifx \href  \undefined \def \href#1#2{\textsf{#2}}\fi
\ifx \doiurl  \undefined \def
  \doiurl#1{\href{http://dx.doi.org/#1}{\textsf{#1}}}\fi
\ifx \betal  \undefined \def \betal{\textit{et al.}}\fi
\ifx \binstitute  \undefined \def \binstitute#1{#1}\fi
\ifx \bctitle  \undefined \def \bctitle#1{#1}\fi
\ifx \beditor  \undefined \def \beditor#1{#1}\fi
\ifx \bpublisher  \undefined \def \bpublisher#1{#1}\fi
\ifx \bbtitle  \undefined \def \bbtitle#1{\textit{#1}}\fi
\ifx \bedition  \undefined \def \bedition#1{#1}\fi
\ifx \bseriesno  \undefined \def \bseriesno#1{\textbf{#1}}\fi
\ifx \blocation  \undefined \def \blocation#1{#1}\fi
\ifx \bsertitle  \undefined \def \bsertitle#1{\textit{#1}}\fi
\ifx \bsnm \undefined \def \bsnm#1{#1}\fi
\ifx \bsuffix \undefined \def \bsuffix#1{#1}\fi
\ifx \bparticle \undefined \def \bparticle#1{#1}\fi
\ifx \barticle \undefined \def \barticle#1{}\fi
\ifx \botherref \undefined \def \botherref#1{}\fi
\ifx \url \undefined \def \url#1{\textsf{#1}}\fi
\ifx \bchapter \undefined \def \bchapter#1{}\fi
\ifx \bbook \undefined \def \bbook#1{}\fi
\ifx \bcomment \undefined \def \bcomment#1{#1}\fi
\ifx \oauthor \undefined \def \oauthor#1{#1}\fi
\ifx \citeauthoryear \undefined \def \citeauthoryear#1{#1}\fi
\def \endbibitem {}
\ifx \bconflocation  \undefined \def \bconflocation#1{#1} \fi

\bibitem[\protect\citeauthoryear{{Alfv\'en}}{1942}]{Alfven42}
\begin{barticle}
\bauthor{\bsnm{{Alfv\'en}}, \binits{H.}}:
\byear{1942},.
\bjtitle{\nat}
\bvolume{150},
\bfpage{405}\,--\,\blpage{406}.
\end{barticle}
\endbibitem

\bibitem[\protect\citeauthoryear{{Allan} and {Manuel}}{1996}]{AM96}
\begin{barticle}
\bauthor{\bsnm{{Allan}}, \binits{W.}},
\bauthor{\bsnm{{Manuel}}, \binits{J.R.}}:
\byear{1996},.
\bjtitle{Annales Geophysicae}
\bvolume{14},
\bfpage{893}\,--\,\blpage{905}.
\end{barticle}
\endbibitem

\bibitem[\protect\citeauthoryear{{Allan}, {Poulter}, and
  {Manuel}}{1991}]{Allan91}
\begin{barticle}
\bauthor{\bsnm{{Allan}}, \binits{W.}},
\bauthor{\bsnm{{Poulter}}, \binits{E.M.}},
\bauthor{\bsnm{{Manuel}}, \binits{J.R.}}:
\byear{1991},.
\bjtitle{\jgr}
\bvolume{96},
\bfpage{11461}.
\end{barticle}
\endbibitem

\bibitem[\protect\citeauthoryear{{Arber} \textit{et~al.}}{2001}]{LAREpaper}
\begin{barticle}
\bauthor{\bsnm{{Arber}}, \binits{T.D.}},
\bauthor{\bsnm{{Longbottom}}, \binits{A.W.}},
\bauthor{\bsnm{{Gerrard}}, \binits{C.L.}},
\bauthor{\bsnm{{Milne}}, \binits{A.M.}}:
\byear{2001},.
\bjtitle{Journal of Computational Physics}
\bvolume{171},
\bfpage{151}\,--\,\blpage{181}.
\end{barticle}
\endbibitem

\bibitem[\protect\citeauthoryear{{Botha} \textit{et~al.}}{2000}]{gert2000}
\begin{barticle}
\bauthor{\bsnm{{Botha}}, \binits{G.J.J.}},
\bauthor{\bsnm{{Arber}}, \binits{T.D.}},
\bauthor{\bsnm{{Nakariakov}}, \binits{V.M.}},
\bauthor{\bsnm{{Keenan}}, \binits{F.P.}}:
\byear{2000},.
\bjtitle{\aap}
\bvolume{363},
\bfpage{1186}\,--\,\blpage{1194}.
\end{barticle}
\endbibitem

\bibitem[\protect\citeauthoryear{Boyd and Sanderson}{2003}]{boydsanderson}
\begin{botherref}
\oauthor{\bsnm{Boyd}, \binits{T.J.M.}},
\oauthor{\bsnm{Sanderson}, \binits{J.J.}}:
2003,,
Cambridge University Press.
\end{botherref}
\endbibitem

\bibitem[\protect\citeauthoryear{{Browning}}{1991}]{Browning91}
\begin{barticle}
\bauthor{\bsnm{{Browning}}, \binits{P.K.}}:
\byear{1991},.
\bjtitle{Plasma Physics and Controlled Fusion}
\bvolume{33},
\bfpage{539}\,--\,\blpage{571}.
\end{barticle}
\endbibitem

\bibitem[\protect\citeauthoryear{{Champeaux}, {Passot}, and
  {Sulem}}{1997}]{Champ97}
\begin{barticle}
\bauthor{\bsnm{{Champeaux}}, \binits{S.}},
\bauthor{\bsnm{{Passot}}, \binits{T.}},
\bauthor{\bsnm{{Sulem}}, \binits{P.L.}}:
\byear{1997},.
\bjtitle{Journal of Plasma Physics}
\bvolume{58},
\bfpage{665}\,--\,\blpage{690}.
\end{barticle}
\endbibitem

\bibitem[\protect\citeauthoryear{Chen}{1984}]{Chen1984}
\begin{botherref}
\oauthor{\bsnm{Chen}, \binits{F.F.}}:
1984,
\textbf{v. 1},
Springer.
\end{botherref}
\endbibitem

\bibitem[\protect\citeauthoryear{{De Moortel}}{2005}]{ineke2005}
\begin{barticle}
\bauthor{\bsnm{{De Moortel}}, \binits{I.}}:
\byear{2005},.
\bjtitle{Royal Society of London Philosophical Transactions Series A}
\bvolume{363},
\bfpage{2743}\,--\,\blpage{2760}.
\end{barticle}
\endbibitem

\bibitem[\protect\citeauthoryear{{De Moortel} and
  {Nakariakov}}{2012}]{ineke2012}
\begin{barticle}
\bauthor{\bsnm{{De Moortel}}, \binits{I.}},
\bauthor{\bsnm{{Nakariakov}}, \binits{V.M.}}:
\byear{2012},.
\bjtitle{Royal Society of London Philosophical Transactions Series A}
\bvolume{370},
\bfpage{3193}\,--\,\blpage{3216}.
\end{barticle}
\endbibitem

\bibitem[\protect\citeauthoryear{{Derby}}{1978}]{Derby78}
\begin{barticle}
\bauthor{\bsnm{{Derby}}, \binits{N.F.} \bsuffix{Jr.}}:
\byear{1978},.
\bjtitle{\apj}
\bvolume{224},
\bfpage{1013}\,--\,\blpage{1016}.
\end{barticle}
\endbibitem

\bibitem[\protect\citeauthoryear{{Dewar}}{1970}]{Dewar70}
\begin{barticle}
\bauthor{\bsnm{{Dewar}}, \binits{R.L.}}:
\byear{1970},.
\bjtitle{Physics of Fluids}
\bvolume{13},
\bfpage{2710}\,--\,\blpage{2720}.
\end{barticle}
\endbibitem

\bibitem[\protect\citeauthoryear{{Falle} and {Hartquist}}{2002}]{FH02}
\begin{barticle}
\bauthor{\bsnm{{Falle}}, \binits{S.A.E.G.}},
\bauthor{\bsnm{{Hartquist}}, \binits{T.W.}}:
\byear{2002},.
\bjtitle{\mnras}
\bvolume{329},
\bfpage{195}\,--\,\blpage{203}.
\end{barticle}
\endbibitem

\bibitem[\protect\citeauthoryear{{Fruit} and
  {Craig}}{2006}]{fruit_nullpointphasemixing}
\begin{barticle}
\bauthor{\bsnm{{Fruit}}, \binits{G.}},
\bauthor{\bsnm{{Craig}}, \binits{I.J.D.}}:
\byear{2006},.
\bjtitle{\aap}
\bvolume{448},
\bfpage{753}\,--\,\blpage{761}.
\end{barticle}
\endbibitem

\bibitem[\protect\citeauthoryear{{Goldstein}}{1978}]{Goldstein78}
\begin{barticle}
\bauthor{\bsnm{{Goldstein}}, \binits{M.L.}}:
\byear{1978},.
\bjtitle{\apj}
\bvolume{219},
\bfpage{700}\,--\,\blpage{704}.
\end{barticle}
\endbibitem

\bibitem[\protect\citeauthoryear{{Goossens}, {Erd{'e}lyi}, and
  {Ruderman}}{2011}]{goosens2011}
\begin{barticle}
\bauthor{\bsnm{{Goossens}}, \binits{M.}},
\bauthor{\bsnm{{Erd{'e}lyi}}, \binits{R.}},
\bauthor{\bsnm{{Ruderman}}, \binits{M.S.}}:
\byear{2011},.
\bjtitle{\ssr}
\bvolume{158},
\bfpage{289}\,--\,\blpage{338}.
\end{barticle}
\endbibitem

\bibitem[\protect\citeauthoryear{{Grimshaw}, {Pelinovsky}, and
  {Pelinovsky}}{2010}]{grimshaw}
\begin{barticle}
\bauthor{\bsnm{{Grimshaw}}, \binits{R.}},
\bauthor{\bsnm{{Pelinovsky}}, \binits{D.}},
\bauthor{\bsnm{{Pelinovsky}}, \binits{E.}}:
\byear{2010},.
\bjtitle{Wave Motion}
\bvolume{99},
\bfpage{496}\,--\,\blpage{507}.
\end{barticle}
\endbibitem

\bibitem[\protect\citeauthoryear{{Heyvaerts} and {Priest}}{1983}]{HP83}
\begin{barticle}
\bauthor{\bsnm{{Heyvaerts}}, \binits{J.}},
\bauthor{\bsnm{{Priest}}, \binits{E.R.}}:
\byear{1983},.
\bjtitle{\aap}
\bvolume{117},
\bfpage{220}\,--\,\blpage{234}.
\end{barticle}
\endbibitem

\bibitem[\protect\citeauthoryear{{McLaughlin}, {De Moortel}, and
  {Hood}}{2011}]{jamesphasemixing2011}
\begin{barticle}
\bauthor{\bsnm{{McLaughlin}}, \binits{J.A.}},
\bauthor{\bsnm{{De Moortel}}, \binits{I.}},
\bauthor{\bsnm{{Hood}}, \binits{A.W.}}:
\byear{2011},.
\bjtitle{\aap}
\bvolume{527},
\bfpage{A149}.
\end{barticle}
\endbibitem

\bibitem[\protect\citeauthoryear{{Nakariakov} and {Verwicter}}{2005}]{NK2005}
\begin{barticle}
\bauthor{\bsnm{{Nakariakov}}, \binits{V.M.}},
\bauthor{\bsnm{{Verwicter}}, \binits{E.O.}}:
\byear{2005},.
\bjtitle{Liv. Rev. Sol. Phys.}
\bvolume{2},
\bfpage{3}.
\end{barticle}
\endbibitem

\bibitem[\protect\citeauthoryear{{Nakariakov}, {Roberts}, and
  {Murawski}}{1997}]{NK97}
\begin{barticle}
\bauthor{\bsnm{{Nakariakov}}, \binits{V.M.}},
\bauthor{\bsnm{{Roberts}}, \binits{B.}},
\bauthor{\bsnm{{Murawski}}, \binits{K.}}:
\byear{1997},.
\bjtitle{\solphys}
\bvolume{175},
\bfpage{93}\,--\,\blpage{105}.
\end{barticle}
\endbibitem

\bibitem[\protect\citeauthoryear{{Narain} and {Ulmschneider}}{1996}]{NU2}
\begin{barticle}
\bauthor{\bsnm{{Narain}}, \binits{U.}},
\bauthor{\bsnm{{Ulmschneider}}, \binits{P.}}:
\byear{1996},.
\bjtitle{\ssr}
\bvolume{75},
\bfpage{453}\,--\,\blpage{509}.
\end{barticle}
\endbibitem

\bibitem[\protect\citeauthoryear{{Parker}}{1991}]{Parker91}
\begin{barticle}
\bauthor{\bsnm{{Parker}}, \binits{E.N.}}:
\byear{1991},.
\bjtitle{\apj}
\bvolume{376},
\bfpage{355}\,--\,\blpage{363}.
\end{barticle}
\endbibitem

\bibitem[\protect\citeauthoryear{{Rankin} \textit{et~al.}}{1994}]{Rankin94}
\begin{barticle}
\bauthor{\bsnm{{Rankin}}, \binits{R.}},
\bauthor{\bsnm{{Frycz}}, \binits{P.}},
\bauthor{\bsnm{{Tikhonchuk}}, \binits{V.T.}},
\bauthor{\bsnm{{Samson}}, \binits{J.C.}}:
\byear{1994},.
\bjtitle{\jgr}
\bvolume{99},
\bfpage{21291}\,--\,\blpage{21302}.
\end{barticle}
\endbibitem

\bibitem[\protect\citeauthoryear{{Ruderman} and
  {Erd{'e}lyi}}{2009}]{Ruderman2009}
\begin{barticle}
\bauthor{\bsnm{{Ruderman}}, \binits{M.S.}},
\bauthor{\bsnm{{Erd{'e}lyi}}, \binits{R.}}:
\byear{2009},.
\bjtitle{\ssr}
\bvolume{149},
\bfpage{199}\,--\,\blpage{228}.
\end{barticle}
\endbibitem

\bibitem[\protect\citeauthoryear{{Stark}, {Musielak}, and
  {Suess}}{1995}]{Stark95}
\begin{botherref}
\oauthor{\bsnm{{Stark}}, \binits{B.A.}},
\oauthor{\bsnm{{Musielak}}, \binits{Z.E.}},
\oauthor{\bsnm{{Suess}}, \binits{S.T.}}:
1995,
In: \textit{Solar Wind Eight},
66.
\end{botherref}
\endbibitem

\bibitem[\protect\citeauthoryear{{Thurgood} and {McLaughlin}}{2012}]{Me2012a}
\begin{barticle}
\bauthor{\bsnm{{Thurgood}}, \binits{J.O.}},
\bauthor{\bsnm{{McLaughlin}}, \binits{J.A.}}:
\byear{2012},.
\bjtitle{\aap}
\bvolume{545},
\bfpage{A9}.
\end{barticle}
\endbibitem

\bibitem[\protect\citeauthoryear{{Tikhonchuk} \textit{et~al.}}{1995}]{Tik95}
\begin{barticle}
\bauthor{\bsnm{{Tikhonchuk}}, \binits{V.T.}},
\bauthor{\bsnm{{Rankin}}, \binits{R.}},
\bauthor{\bsnm{{Frycz}}, \binits{P.}},
\bauthor{\bsnm{{Samson}}, \binits{J.C.}}:
\byear{1995},.
\bjtitle{Phys. Plasmas}
\bvolume{2},
\bfpage{501}\,--\,\blpage{515}.
\end{barticle}
\endbibitem

\bibitem[\protect\citeauthoryear{{Ulmschneider} and {Narain}}{1990}]{NU1}
\begin{botherref}
\oauthor{\bsnm{{Ulmschneider}}, \binits{P.}},
\oauthor{\bsnm{{Narain}}, \binits{U.}}:
1990,
In: {Priest}, E.R., {Krishan}, V. (eds.)
\textit{Basic Plasma Processes on the Sun},
\textit{IAU Symposium}
\textbf{142},
97.
\end{botherref}
\endbibitem

\bibitem[\protect\citeauthoryear{{Verwichte}}{1999}]{ErwinThesis}
\begin{botherref}
\oauthor{\bsnm{{Verwichte}}, \binits{E.}}:
1999,.
PhD thesis,
The Open University.
\end{botherref}
\endbibitem

\bibitem[\protect\citeauthoryear{{Verwichte}, {Nakariakov}, and
  {Longbottom}}{1999}]{Erwin99}
\begin{barticle}
\bauthor{\bsnm{{Verwichte}}, \binits{E.}},
\bauthor{\bsnm{{Nakariakov}}, \binits{V.M.}},
\bauthor{\bsnm{{Longbottom}}, \binits{A.W.}}:
\byear{1999},.
\bjtitle{Journal of Plasma Physics}
\bvolume{62},
\bfpage{219}\,--\,\blpage{232}.
\end{barticle}
\endbibitem

\bibitem[\protect\citeauthoryear{{Webb} \textit{et~al.}}{2001}]{Webb01}
\begin{barticle}
\bauthor{\bsnm{{Webb}}, \binits{G.M.}},
\bauthor{\bsnm{{Zakharian}}, \binits{A.R.}},
\bauthor{\bsnm{{Brio}}, \binits{M.}},
\bauthor{\bsnm{{Zank}}, \binits{G.P.}}:
\byear{2001},.
\bjtitle{Journal of Plasma Physics}
\bvolume{66},
\bfpage{167}\,--\,\blpage{212}.
\end{barticle}
\endbibitem

\bibitem[\protect\citeauthoryear{{Webb} \textit{et~al.}}{2005}]{webb05pt2}
\begin{barticle}
\bauthor{\bsnm{{Webb}}, \binits{G.M.}},
\bauthor{\bsnm{{Zank}}, \binits{G.P.}},
\bauthor{\bsnm{{Kaghashvili}}, \binits{E.K.}},
\bauthor{\bsnm{{Ratkiewicz}}, \binits{R.E.}}:
\byear{2005},.
\bjtitle{Journal of Plasma Physics}
\bvolume{71},
\bfpage{811}\,--\,\blpage{857}.
\end{barticle}
\endbibitem

\end{thebibliography}


%

\end{article} 

\end{document}